\documentclass[aps,prb,twocolumn,10pt,superscriptaddress,showpacs]{revtex4-2}
\usepackage{amsmath,amssymb,graphics,epsfig,epstopdf,color,verbatim,ulem,braket,tabularx,multirow,array,diagbox,colortbl,hhline}
\usepackage[colorlinks,linkcolor=blue,citecolor=blue,urlcolor=blue]{hyperref}
\usepackage{physics}
\usepackage{bm}
\usepackage{blindtext}
\graphicspath{{Figures/}}

\begin{document}

\title{Spin-triplet pairing instability in a two-dimensional repulsive Hubbard model}

\author{Xing-Can Liu}
\affiliation{School of physics, Harbin Institute of Technology, Harbin 150001, China}
\affiliation{Institute of Modern Physics, Northwest University, Xi'an 710127, China}

\author{Yu-Feng Song}
\affiliation{Institute of Modern Physics, Northwest University, Xi'an 710127, China}
\affiliation{Hefei National Laboratory for Physical Sciences at Microscale and Department of Modern Physics, University of Science and Technology of China, Hefei, Anhui 230026, China}

\author{Yuan-Yao He}
\email{heyuanyao@nwu.edu.cn}
\affiliation{Institute of Modern Physics, Northwest University, Xi'an 710127, China}
\affiliation{Shaanxi Key Laboratory for Theoretical Physics Frontiers, Xi'an 710127, China}
\affiliation{Peng Huanwu Center for Fundamental Theory, Xian 710127, China}
\affiliation{Hefei National Laboratory, Hefei 230088, China}

\author{Tao Ying}
\email{taoying86@hit.edu.cn}
\affiliation{School of physics, Harbin Institute of Technology, Harbin 150001, China}

\author{Xueru Zhang}
\email{xrzhang@hit.edu.cn}
\affiliation{School of physics, Harbin Institute of Technology, Harbin 150001, China}

\begin{abstract}
The search for superconductivity with unconventional pairing symmetry has been a central focus in the study of strongly correlated electron systems. In this work, we report a numerically exact study of the spin-triplet pairing in a two-dimensional Hubbard model with repulsive interactions, employing Determinant Quantum Monte Carlo method. The model includes next-nearest-neighbor and third-nearest-neighbor hopping terms, and maintains spin balance. By tuning the fermion filling close to a type-II van Hove singularity (vHs) in the model, we numerically investigate the ordering tendencies of several possible pairing channels with different symmetries. Our numerical results provide clear evidence for the spin-triplet $p$-wave pairing instability approaching low temperatures, as revealed by the vertex contribution to the pairing susceptibility. This signature becomes increasingly pronounced as the interaction strength increases in the weak to intermediate regime. We further find that, near the type-II vHs, the dominant spin-spin correlations in the system are ferromagnetic, suggesting its close relation to the spin-triplet pairing instability. Our findings offer a reliable approach to realize the spin-triplet $p$-wave superfluidity in the repulsive Hubbard model, from an unbiased numerical perspective. 
\end{abstract}

\date{\today}

\maketitle

\section{Introduction}
\label{sec:intro}

Superconductivity is one of the long-standing interests in condensed matter physics. The key to understanding its microscopic mechanism lies in identifying the pairing symmetry of the electron pairs (or Cooper pairs). In conventional metallic and alloy superconductors (SCs), the Bardeen-Cooper-Schrieffer (BCS) theory attributes the pairing interaction between electrons to phonon mediation, which typically leads to the spin-singlet pairing with $s$-wave symmetry~\cite{Bardeen1957}. The subsequent discovery of unconventional SCs, represented by the cuprates hosting high-temperature superconductivity~\cite{bednorz1986,Dagotto1994,Tsuei2000}, has revealed fundamentally different electron pairing symmetries. One example is the spin-singlet $d$-wave pairing observed in cuprates. An even more exotic case is the spin-triplet odd-parity pairing, in which the Cooper pairs form a spin state with total spin $S=1$. Such a pairing state was first confirmed between the fermionic atoms in the superfluid phases of $^3$He~\cite{Osheroff1972,Anderson1973}. In realistic materials, the experimental studies have found several promising candidates, including ${\rm Sr_2RuO_4}$~\cite{Mackenzie2003,Maeno2024}, ${\rm UTe_2}$~\cite{Sheng2019,Aoki2022}, and ${\rm UPt_3}$~\cite{Joynt2002,Avers2020}. These materials have also been proposed as potential hosts of topological superconductivity~\cite{Ivanov2001,Kallin2012}, which could enable applications in quantum computation~\cite{Sarma2006,Nayak2008}. 

Besides concrete examples of correlated electron materials, an alternative route to realize spin-triplet superconductivity in two-dimensional (2D) systems, as firstly suggested in Ref.~\cite{Yao2015}, is to tune the electron filling close to a type-II van Hove singularity (vHs). This proposal is based on the odd-parity gap function $\Delta_t(\mathbf{k})$ for triplet pairing, as $\Delta_t(-\mathbf{k})=-\Delta_t(\mathbf{k})$. As a result, $\Delta_t(\mathbf{k})$ vanishes at type-I vHs, corresponding to saddle points in noninteracting energy dispersion located at time-reversal invariant momenta (TRIM), i.e., $\mathbf{K}=-\mathbf{K}$. This leads to a general suppression of triplet pairing at type-I vHs. In contrast, at type-II vHs where the saddle points are located away from TRIM, this suppression is lifted, allowing spin-triplet pairing to potentially emerge. Building on the insight, further renormalization group (RG) analysis in Ref.~\cite{Yao2015} demonstrated that triplet pairing is most favorable at type-II vHs in the weak-coupling limit with repulsive interactions. This prediction was subsequently confirmed by dynamical mean-field theory (DMFT) simulations on a square-lattice Hubbard model~\cite{Meng2015}. Moreover, these results have implications for several 2D materials, including layered superconductors LaO$_{1-x}$F$_x$BiS$_2$~\cite{Mizuguchi2012,Usui2012,Yang2013}, the doped BC$_3$~\cite{Chen2015,Wang2018}, and van der Waals material NiPS$_3$~\cite{Chittari2016,Li2021}. The ab initio calculations for these materials~\cite{Usui2012,Chen2015,Chittari2016} revealed the existence of the type-II vHs, and further random phase approximation (RPA)~\cite{Chen2015,Li2021} as well as functional RG~\cite{Yang2013,Wang2018} studies also suggested the possibility of spin-triplet superconductivity in these systems. {  Nevertheless, all these theoretical studies rely on certain approximations built into the methods (including RG, DMFT, RPA, and functional RG). It is therefore highly desirable to address the above problem using a nonperturbative and numerically exact method, serving as an independent and complementary examination.}

%These theoretical studies offer compelling evidence for this scenario. However, they are built upon approximations whose controlled application in strongly correlated regimes can be challenging. It is therefore highly desirable to address this problem with a method that is non-perturbative and numerically exact on finite lattices. 

In this work, we aim to revisit the spin-triplet pairing instability near the type-II vHs in a numerically exact manner, employing Determinant Quantum Monte Carlo (DQMC) method. We adopt the square-lattice Hubbard model with repulsive interaction, as previously studied in Refs.~\cite{Yao2015,Meng2015}. In our numerical calculations, all relevant electron pairing channels are treated on equal footing, including the spin-singlet $d_{x^2-y^2}$ and $d_{xy}$ channels, as well as the spin-triplet $p_x$+$ip_y$ and $p_{xy}$ channels. We focus on corresponding static pairing susceptibilities and their vertex contributions. Our results clearly demonstrate that triplet pairing dominates near the type-II vHs in the weak to intermediate interaction regime. Meanwhile, we observe ferromagnetic spin correlations, which may serve as the underlying mechanism for the emergence of triplet pairing. We also estimate the transition temperature for the spin-triplet superconductivity in the system via the scaling behavior of the pairing susceptibility.

The remainder of the paper is organized as follows. In Sec.~\ref{sec:Modelmethod}, we introduce the square-lattice Hubbard model and the type-II vHs in the model, briefly discuss the DQMC method, and outline the physical observables for the relevant pairing channels. In Sec.~\ref{sec:Results}, we concentrate on our numerical results for various pairing channels as well as for the spin channel. Finally, we present a summary and discussion of our work in Sec.~\ref{sec:Summary}. The Appendix contain the derivations for the pairing susceptibility in free fermion systems.

\section{Model, method, and physical observables}
\label{sec:Modelmethod}

\subsection{The square-lattice Hubbard model}
\label{sec:TheModel}
We study the 2D Fermi-Hubbard model on a square lattice, described by the following Hamiltonian:
\begin{equation}\begin{aligned}
\label{eq:2DHamlt}
\hat{H} = &\sum_{\mathbf{ij},\sigma} t_{\mathbf{ij}} c^+_{\mathbf{i}\sigma} c^{}_{\mathbf{j}\sigma} 
+ \mu\sum_{\mathbf{i},\sigma}\hat{n}_{\mathbf{i}\sigma} \\
&\hspace{1.0cm} + U\sum_{\mathbf{i}}\Big(\hat{n}_{\mathbf{i}\uparrow}\hat{n}_{\mathbf{i}\downarrow}-\frac{\hat{n}_{\mathbf{i}\uparrow}+\hat{n}_{\mathbf{i}\downarrow}}{2}\Big),
\end{aligned}\end{equation}
where $\hat{n}_{\mathbf{i}\sigma}=c^+_{\mathbf{i}\sigma} c^{}_{\mathbf{i}\sigma}$ denotes the density operator, with $\sigma$ ($=\uparrow,\downarrow$) as the spin of electrons and $\mathbf{i}=(i_x,i_y)$ as coordinates of the lattice site. In the noninteracting part, we set $t_{\mathbf{ij}}=t_1,t_2,t_3$ as the strengths for first-, second-, and third-nearest-neighbor hoppings, respectively. We also assume $|t_2|<t$ and $t_3<0$. This yields the energy dispersion $\varepsilon_{\mathbf{k}}=2t_1(\cos k_x+\cos k_y) + 4t_2\cos k_x\cos k_y+2t_3[\cos(2k_x)+\cos(2k_y)]$, where $k_x,k_y$ are the momentum defined in units of $2\pi/L$ with linear system size $L$. In this work, we adopt $t_1=-t$ and set $t=1$ as the energy unit, and focus on repulsive interaction $U>0$. The chemical potential term $\mu>0$ represents the hole doping. The electron filling is defined as $n = N_e / N_s$, where $N_e$ is the number of electrons, and $N_s = L^2$ is the number of lattice sites. 

\begin{figure}[t]
\centering
\includegraphics[width=0.75\columnwidth]{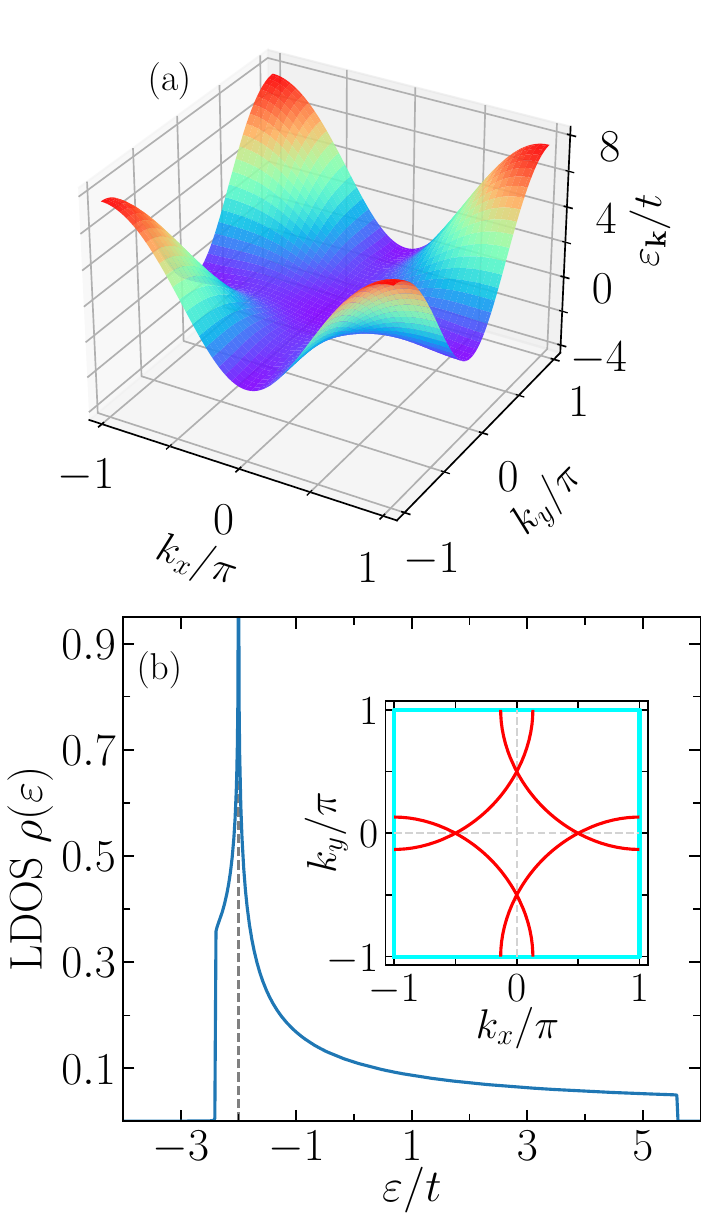}
\caption{The plots of (a) energy dispersion $\varepsilon_{\mathbf{k}}/t$, and (b) local density of states (LDOS) $\rho(\varepsilon)$, for the noninteracting part of the model~(\ref{eq:2DHamlt}), with $t_2/t=+0.5$ and $t_3/t=-0.1$. The Fermi surface for $\mu=(t_1+2t_2)^2/(4t_3)-2t_1=2t$ is plotted in inset of panel (b). The saddle points located at $\mathbf{K}_1=(\pm\pi/2,0)$ and $\mathbf{K}_2=(0,\pm\pi/2)$ induce the type-II vHs at $\varepsilon=-2t$ (vertical green dashed line) in (b), and the corresponding electron filling is $n_{\rm v}=0.3757$. }
\label{fig:fermisurf}
\end{figure}
 
For a 2D noninteracting system, the local density of states (LDOS) $\rho(\varepsilon)$ must host logarithmically divergent vHs~\cite{Furukawa1998,Honerkamp2001}, which is contributed by saddle points in the energy dispersion. For the model~(\ref{eq:2DHamlt}) with $U=0$ and $|t_1+2t_2|>-4t_3$, such vHs appears at $\varepsilon_{\rm I}=4(t_3-t_2)$, and originates to saddle points in the $\varepsilon_{\mathbf{k}}$ dispersion located at the TRIM $\mathbf{X}_1=(\pm\pi,0)$ and $\mathbf{X}_2=(0,\pm\pi)$. This case is referred as type-I vHs. For $|t_1+2t_2|<-4t_3$, saddle points in $\varepsilon_{\mathbf{k}}$ at momenta (depending on $t_2$ and $t_3$) away from all TRIM induce the type-II vHs at $\varepsilon_{\rm II}=-(t_1+2t_2)^2/(4t_3)+2t_1$. By tuning the chemical potential to $\mu=-\varepsilon_{\rm I}$ or $\mu=-\varepsilon_{\rm II}$, the vHs appears on the Fermi surface (FS), leading to a log-square divergence in the superconducting susceptibilities for both spin-singlet and triplet pairings~\cite{Furukawa1998,Yao2015,Xie2025}. As mentioned in Introduction, triplet pairing is generally disfavored at type-I vHs due to symmetry constraint, whereas it can be stabilized at type-II vHs. Furthermore, in the weakly interacting limit, the above FS physics near vHs governs the low-energy behaviors of the model~(\ref{eq:2DHamlt}), and serves as a natural starting point for theoretical analysis~\cite{Yao2015,Meng2015}.

We choose a representative parameter set, $t_2/t=+0.5$ and $t_3/t=-0.1$, and plot the energy dispersion $\varepsilon_{\mathbf{k}}/t$ and its corresponding LDOS $\rho(\varepsilon)$ for the model~(\ref{eq:2DHamlt}) in Fig.~\ref{fig:fermisurf}. Here, LDOS is computed via $\rho(\varepsilon)=N_s^{-1}\sum_{\mathbf{k}}\delta(\varepsilon-\varepsilon_{\mathbf{k}})$ for both spin-up and spin-down channels. For this case, the saddle points in $\varepsilon_{\mathbf{k}}$ locate at $\mathbf{K}_1=(\pm\pi/2,0)$ and $\mathbf{K}_2=(0,\pm\pi/2)$, all of which are not TRIM. The corresponding type-II vHs induced by the saddle points appears at $\varepsilon_{\rm II}=-2t$ in LDOS, yielding an electron filling $n_{\rm v}=0.3757$ at the vHs. In this work, we concentrate on this parameter set, and perform the numerical calculations around type-II vHs via tuning $\mu$ to control the electron filling.

\subsection{DQMC method}
\label{sec:DQMCmethod}

We apply the finite-temperature DQMC method~\cite{Blankenbecler1981,Hirsch1983,White1989,Scalettar1991,McDaniel2017,Sun2024,Yuanyao2019,Yuanyao2019L,Assaad2008,Chang2015} to numerically solve the lattice model in Eq.~(\ref{eq:2DHamlt}). The key ingredient in DQMC is the Hubbard-Stratonovich (HS) transformation~\cite{Hirsch1983,Assaad1998,Shihao2013,WangDa2014}, which is used to decouple the two-body interactions in correlated fermion systems into free fermions coupled to fluctuating auxiliary fields. Fermionic observables are then evaluated via Markov-Chain Monte Carlo (MCMC) sampling over auxiliary-field configurations. In the following, we briefly outline the DQMC framework and the calculations for physical observables.

The DQMC framework for the model~(\ref{eq:2DHamlt}) starts with the partition function as $Z={\rm Tr}(e^{-\beta\hat{H}})$. In our calculations, it proceeds with the imaginary-time discretization $\beta=M\Delta\tau$, Trotter-Suzuki decomposition (such as the formula $e^{-\Delta\tau\hat{H}}=e^{-\frac{\Delta\tau}{2}\hat{H}_0}e^{-\Delta\tau\hat{H}_U}e^{-\frac{\Delta\tau}{2}\hat{H}_0}+\mathcal{O}[(\Delta\tau)^3]$ where $\hat{H}_0$ and $\hat{H}_U$ represent the kinetic and interaction parts, respectively), and the HS transformation into spin-$\hat{s}^z$ channel reading
\begin{equation}\begin{aligned}
\label{eq:HSspinDecomp}
e^{-\Delta\tau U \big(\hat{n}_{\mathbf{i}\uparrow}\hat{n}_{\mathbf{i}\downarrow}-\frac{\hat{n}_{\mathbf{i}\uparrow}+\hat{n}_{\mathbf{i}\downarrow}}{2}\big) }
= \frac{1}{2}\sum_{x_{\mathbf{i}}=\pm1}e^{\gamma x_{\mathbf{i}}(\hat{n}_{\mathbf{i}\uparrow}-\hat{n}_{\mathbf{i}\downarrow})},
\end{aligned}\end{equation}
with the auxiliary field $x_{\mathbf{i}}$ and the coupling coefficient $\gamma=\cosh^{-1}(e^{+\Delta\tau U/2})$ for $U>0$ case. After applying these procedures to all $e^{-\Delta\tau\hat{H}}$ operators in $Z$, we can arrive at $Z\simeq \sum_{\mathbf{X}}P(\mathbf{X}){\rm Tr}(\hat{B}_M\cdots\hat{B}_2\hat{B}_1)$, with the probability $P(\mathbf{X})=2^{-MN_s}$ and the complete auxiliary-field configuration $\mathbf{X}=(\mathbf{x}_{M},\cdots,\mathbf{x}_{\ell},\cdots,\mathbf{x}_{1})$ with $\mathbf{x}_{\ell}=(x_{\ell,1},x_{\ell,2},\cdots,x_{\ell,N_s})$. The single-particle propagator $\hat{B}_{\ell}$ takes the form $\hat{B}_{\ell}=e^{-\Delta\tau\hat{H}_0/2}\hat{B}_I(\mathbf{x}_{\ell})e^{-\Delta\tau\hat{H}_0/2}$ with $\hat{B}_I(\mathbf{x}_{\ell})=\prod_{\mathbf{i}}e^{\gamma x_{\ell,\mathbf{i}}(\hat{n}_{\mathbf{i}\uparrow}-\hat{n}_{\mathbf{i}\downarrow})}$. The ``$\simeq$'' symbol indicates the Trotter error. Then the trace in $Z$ can be evaluated, which yields $Z\simeq\sum_{\mathbf{X}}W(\mathbf{X})$ with $W(\mathbf{X})$ as the configuration weight, given by~\cite{Assaad2008}
\begin{equation}\begin{aligned}
\label{eq:ConfgWeight}
W(\mathbf{X})
= P(\mathbf{X})\prod_{\sigma=\uparrow,\downarrow}\det(\mathbf{B}_{M}^{\sigma}\cdots\mathbf{B}_{\ell}^{\sigma}\cdots\mathbf{B}_{2}^{\sigma}\mathbf{B}_{1}^{\sigma}),
\end{aligned}\end{equation}
where $\mathbf{B}_{\ell}^{\sigma}=e^{-\Delta\tau\mathbf{H}_0^{\sigma}/2}e^{\mathbf{H}_U^{\sigma}(\mathbf{x}_{\ell})}e^{-\Delta\tau\mathbf{H}_0^{\sigma}/2}$ is the $N_s\times N_s$ propagator matrix, with $\mathbf{H}_0^{\sigma}$ as the hopping matrix in $\hat{H}_0$ and $\mathbf{H}_U^{\sigma}(\mathbf{x}_{\ell})$ as the representation matrix for $\hat{H}_U$ after the HS transformation in Eq.~(\ref{eq:HSspinDecomp}). The spin-decoupled situation for the model~(\ref{eq:2DHamlt}) within Eq.~(\ref{eq:HSspinDecomp}) is manifested in above expression of $W(\mathbf{X})$ in Eq.~(\ref{eq:ConfgWeight}). Then the thermal expectation of a physical observable $\hat{O}$ of fermions can then be expressed as
\begin{equation}\begin{aligned}
\label{eq:PhyObs}
\langle\hat{O}\rangle
=\frac{{\rm Tr}(e^{-\beta\hat{H}}\hat{O})}{Z}
=\frac{\sum_{\mathbf{X}}W(\mathbf{X})O(\mathbf{X})}{\sum_{\mathbf{X}^{\prime}}W(\mathbf{X}^{\prime})}
=\sum_{\mathbf{X}}\omega(\mathbf{X})O(\mathbf{X})
\end{aligned}\end{equation}
where $O(\mathbf{X})=\langle\hat{O}\rangle_{\mathbf{X}}$ is the measurement of $\hat{O}$ within the configuration $\mathbf{X}$. In DQMC method, the summation (or integral) over $\mathbf{X}$ in Eq.~(\ref{eq:PhyObs}) is computed via MCMC sampling using $\omega(\mathbf{X})=W(\mathbf{X})/\sum_{\mathbf{X}^{\prime}}W(\mathbf{X}^{\prime})$ as the probability density function (PDF). The DQMC simulation for the model~(\ref{eq:2DHamlt}) suffers from the minus sign problem~\cite{Loh1990,Troyer2005,Iglovikov2015}, and a reweighting technique needs to be used to reformulate Eq.~(\ref{eq:PhyObs}) to
\begin{equation}\begin{aligned}
\label{eq:PhyObsSign}
\langle\hat{O}\rangle
=\frac{\sum_{\mathbf{X}}\omega^{\prime}(\mathbf{X}){\rm Sgn}(\mathbf{X})O(\mathbf{X})}{\sum_{\mathbf{X}^{\prime}}\omega^{\prime}(\mathbf{X}^{\prime}){\rm Sgn}(\mathbf{X}^{\prime})}
=\frac{\langle\mathcal{S}\hat{O}\rangle}{\langle\mathcal{S}\rangle},
\end{aligned}\end{equation}
with $\omega^{\prime}(\mathbf{X})=|W(\mathbf{X})|/\sum_{\mathbf{X}^{\prime}}|W(\mathbf{X}^{\prime})|$ as the new nonnegative PDF used for the sampling, ${\rm Sgn}(\mathbf{X})={\rm Sign}[W(\mathbf{X})]$ as the sign of the weight, and $\langle\mathcal{S}\rangle=\sum_{\mathbf{X}^{\prime}}\omega^{\prime}(\mathbf{X}^{\prime}){\rm Sgn}(\mathbf{X}^{\prime})$ denoting the sign average.

In this work, we adopt the susceptibility to characterize the pairing instability in the model~(\ref{eq:2DHamlt}). Thus, we focus on the dynamical correlation functions in our DQMC simulations. Such calculations rely on two kinds of time-dependent single-particle Green's functions, represented by $\hat{O}_1=c_{\mathbf{i}\sigma}^{}(\tau)c_{\mathbf{j}\sigma}^+(0)$ and $\hat{O}_2=c_{\mathbf{i}\sigma}^+(\tau)c_{\mathbf{j}\sigma}^{}(0)$ operators, respectively, in Eqs.~(\ref{eq:PhyObs}) and (\ref{eq:PhyObsSign}) [with $\hat{A}(\tau)=e^{\tau\hat{H}}\hat{A}e^{-\tau\hat{H}}$]. The corresponding measurements $O_1(\mathbf{X})$ and $O_2(\mathbf{X})$ are elements of the Green's function matrices $\mathbf{G}_1^{\sigma}(\tau,0)$ and $\mathbf{G}_2^{\sigma}(\tau,0)$, which can be expressed with $\mathbf{B}_{\ell}^{\sigma}$ matrix (see Refs.~\cite{White1989,Assaad2008} for more details and derivations). The measurement of any two-body correlation function within $\mathbf{X}$ involving four creation or annihilation operators can be evaluated via Wick decomposition~\cite{Assaad2008}. Our simulations also incorporate the fast Fourier transform between the real and momentum space~\cite{Yuanyao2019L} during the propagation, along with delayed update~\cite{Sun2024} and global update~\cite{Scalettar1991} of the auxiliary-field configurations. Together, these techniques enhance the efficiency of DQMC simulations.

\subsection{Physical observables}
\label{sec:DQMCObs}

In our DQMC calculations, we concentrate on the susceptibilities of various pairing channels, including the spin-singlet $d_{x^2-y^2}$ and $d_{xy}$ pairing as well as the spin-triplet $p_x$+$ip_y$ and $p_{xy}$ pairing. {  We select these pairing channels because they have been identified as the most relevant ones for the square-lattice Hubbard model~\cite{White1989}, particularly in previous RG analyses~\cite{Yao2015} and DMFT calculations~\cite{Meng2015}.} These pairing patterns can be unified into the following real-space ($\mathbf{r}$-space) pairing operator as
\begin{equation}\begin{aligned}
\label{eq:PairOp1}
\hat{\Delta}_{\mathbf{i}} = \frac{1}{4}\sum_{\boldsymbol{\delta}}\big[ f(\boldsymbol{\delta})\hat{\Lambda}_{\mathbf{i},\boldsymbol{\delta}} + f^{\star}(\boldsymbol{\delta})\hat{\Lambda}_{\mathbf{i},\boldsymbol{\delta}}^{+}\big],
\end{aligned}\end{equation}
where $\boldsymbol{\delta}$ represents the nearest or next-nearest neighbors, and $f(\boldsymbol{\delta})$ is the $\mathbf{r}$-space form factor for the pairing. $\hat{\Lambda}_{\mathbf{i},\boldsymbol{\delta}}$ is the two-site pairing operator given by
\begin{equation}\begin{aligned}
\label{eq:PairOp2}
\hat{\Lambda}_{\mathbf{i},\boldsymbol{\delta}}
= c_{\mathbf{i}\uparrow}^{}c_{\mathbf{i}+\boldsymbol{\delta}\downarrow}^{} \pm c_{\mathbf{i}\downarrow}^{}c_{\mathbf{i}+\boldsymbol{\delta}\uparrow}^{}
\end{aligned}\end{equation}
where ``$-$'' and ``$+$'' represent the spin-singlet and spin-triplet pairing, respectively. Here, we only consider the triplet component with angular momentum $l=0$ due to the its degeneracy with the $l=\pm1$ components. 

\begin{figure}[t]
\centering
\includegraphics[width=0.70\columnwidth]{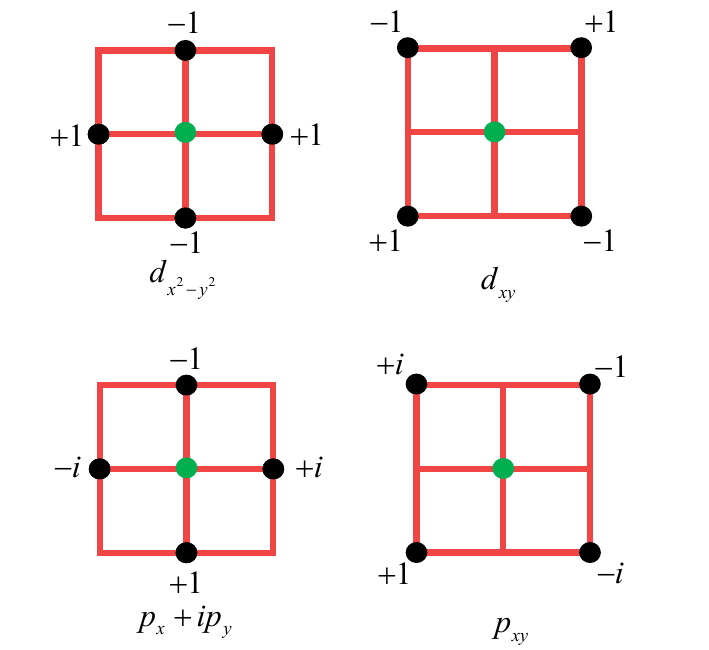}
\caption{Illustration of the $\mathbf{r}$-space form factor $f(\boldsymbol{\delta})$ for the spin-singlet $d_{x^2-y^2}$ and $d_{xy}$ pairing, as well as the spin-triplet $p_x$+$ip_y$ and $p_{xy}$ pairing. The center point (green circle) denotes the $\mathbf{i}=(i_x,i_y)$ site, and the surrounding points (black circles) represent the $\mathbf{i}+\boldsymbol{\delta}$ site in $\hat{\Lambda}_{\mathbf{i},\boldsymbol{\delta}}$ [see Eq.~(\ref{eq:PairOp2})].} 
\label{fig:pairtype}
\end{figure}

The $\mathbf{r}$-space form factor $f(\boldsymbol{\delta})$ for the above four types of pairing patterns~\cite{Zhangkai2025} take the forms
\begin{equation}\begin{aligned}
\label{eq:RspFormFac}
&{f_{{d_{{x^2} - {y^2}}}}}\left( \boldsymbol{\delta} \right) = \left\{ \begin{array}{ll}
{+1}, & {\boldsymbol{\delta} = \left( { \pm 1,0} \right)} \\
{-1}, & {\boldsymbol{\delta} = \left( {0, \pm 1} \right)}
\end{array} \right., \\
&{f_{{p_x} + i{p_y}}}\left( \boldsymbol{\delta} \right) = \left\{ \begin{array}{ll}
{\pm i}, & {\boldsymbol{\delta} = \left( { \pm 1,0} \right)} \\
{\pm 1}, & {\boldsymbol{\delta} = \left( {0, \mp 1} \right)}
\end{array} \right., \\
&{f_{{d_{xy}}}}\left( \boldsymbol{\delta} \right) = \left\{ \begin{array}{ll}
{+1}, & {\boldsymbol{\delta} = \left( {+1, +1} \right), \left( {-1, -1} \right)} \\
{-1}, & {\boldsymbol{\delta} = \left( {+1, -1} \right), \left( {-1, +1} \right)}
\end{array} \right., \\
&{f_{{p_{xy}}}}\left( \boldsymbol{\delta} \right) = \left\{ \begin{array}{ll}
{-1}, & {\boldsymbol{\delta} = \left( {+1, +1} \right)} \\
{+1}, & {\boldsymbol{\delta} = \left( {-1, -1} \right)} \\
{+i}, & {\boldsymbol{\delta} = \left( {+1, -1} \right)} \\
{-i}, & {\boldsymbol{\delta} = \left( {-1, +1} \right)}.
\end{array} \right.
\end{aligned}\end{equation}
The schematic demonstration for these $f(\boldsymbol{\delta})$ are shown in Fig.~\ref{fig:pairtype}. The spin-singlet $d_{x^2-y^2}$-wave and spin-triplet $p_x$+$ip_y$-wave channels are of the nearest-neighbor (NN) pairing, while spin-singlet $d_{xy}$-wave and spin-triplet $p_{xy}$-wave pairings consist of next-nearest-neighbor (NNN) electrons. Here, we only consider the $p_x$+$ip_y$-wave pairing since it is degenerate with the $p_x$$-$$ip_y$-wave pairing. The momentum-space form factor can be obtained as $f(\mathbf{k})=\sum_{\boldsymbol{\delta}}f(\boldsymbol{\delta})e^{+i\mathbf{k}\cdot\boldsymbol{\delta}}/2$. 

%{  All susceptibilities are measured without imposing any explicit symmetry breaking; candidate channels are probed via form factors while the auxiliary fields are sampled unbiasedly within DQMC.}

\begin{figure}[t]
\centering
\includegraphics[width=0.85\linewidth]{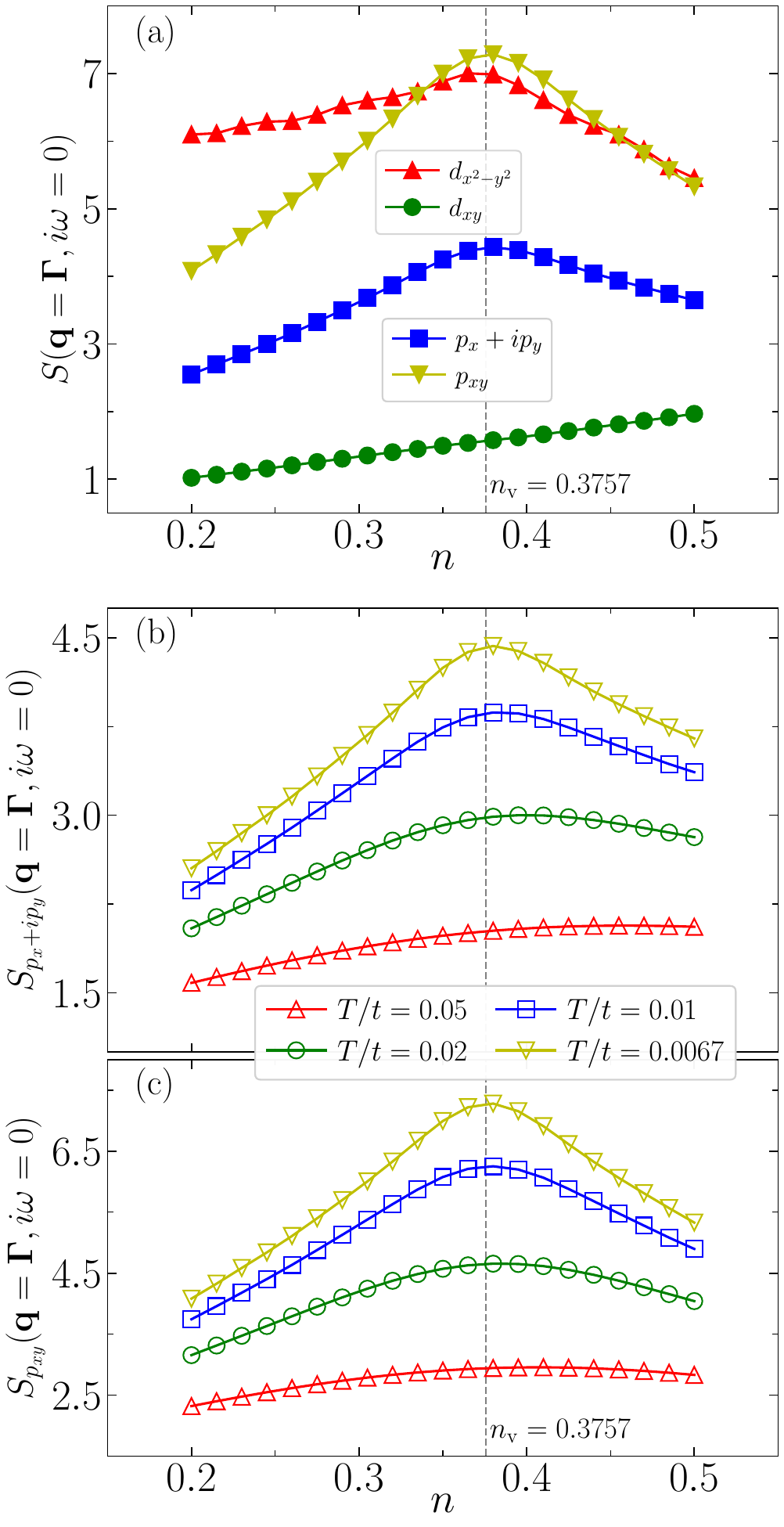}
\caption{Static pairing susceptibilities $S(\mathbf{q}=\boldsymbol{\Gamma},i\omega=0)$ as a function of electron filling $n$ for the noninteracting system with $t_2/t=+0.5,t_3/t=-0.1$. Panel (a) plots the results for $d_{x^2-y^2}$, $d_{xy}$, $p_x$+$ip_y$, and $p_{xy}$ pairing channels at $T/t=0.0067$, while panels (b) and (c) are the results for $p_x$+$ip_y$ and $p_{xy}$ pairings, respectively, with four different temperatures $T/t=0.05,0.02,0.01,0.0067$. The vertical gray dashed lines mark the electron filling $n_{\rm v}=0.3757$ at type-II vHs.}
\label{fig:U0PairSuscep}
\end{figure}

With the above definition for $\hat{\Delta}_{\mathbf{i}}$, the dynamic pairing correlation function can be computed as
\begin{equation}\begin{aligned}
\label{eq:DynPairCrFt}
S(\mathbf{q},\tau) 
= \frac{1}{N_s}\sum_{\mathbf{ij}} e^{-i\mathbf{q}\cdot(\mathbf{R}_{\mathbf{i}} - \mathbf{R}_{\mathbf{j}})}\big\langle\hat{\Delta}_{\mathbf{i}}(\tau)\hat{\Delta}_{\mathbf{j}}(0)\big\rangle,
\end{aligned}\end{equation}
which yields the pairing susceptibility
\begin{equation}\begin{aligned}
\label{eq:PairSuscep}
S(\mathbf{q},i\omega_m) 
= \int_{0}^{\beta} e^{i\omega_m\tau} S(\mathbf{q},\tau) d\tau,
\end{aligned}\end{equation}
with $\omega_m=2m\pi/\beta$ ($m$ is integer) as the Matsubara frequency for bosons. Here we focus on the static pairing susceptibility $S(\mathbf{q},i\omega=0)$ to characterize the ordering tendencies for various pairing orders in the model~(\ref{eq:2DHamlt}) at finite temperature. The ordering vector for the pairing orders involved here is $\mathbf{q}=\boldsymbol{\Gamma}$. We present the analytical formula of $S(\mathbf{q},i\omega=0)$ for the noninteracting system ($U=0$) in Appendix~\ref{sec:AppendixA}, and plot the corresponding numerical results of $S(\boldsymbol{\Gamma},i\omega=0)$ for $t_2/t=+0.5$ and $t_3/t=-0.1$ in Fig.~\ref{fig:U0PairSuscep}. For $d_{x^2-y^2}$, $p_x$+$ip_y$, and $p_{xy}$ pairing channels, $S(\boldsymbol{\Gamma},i\omega=0)$ shows peaks around the type-II vHs [see Fig.~\ref{fig:U0PairSuscep}(a)], which is induced by the divergent LDOS. This clearly illustrates the competition between singlet and triplet pairings. For $p_x$+$ip_y$-wave and $p_{xy}$-wave pairings, the peaks in $S(\boldsymbol{\Gamma},i\omega=0)$ become more pronounced upon cooling, and the peak value should possess the diverging behavior as $\propto -\ln T$~\cite{Xie2025}.

We also calculate the magnetic properties of the system, characterized by the spin susceptibility 
\begin{equation}\begin{aligned}
\label{eq:SpinZZSuscep}
S^{zz}(\mathbf{q},i\omega_m) 
= \int_{0}^{\beta} e^{i\omega_m\tau} S^{zz}(\mathbf{q},\tau) d\tau,
\end{aligned}\end{equation}
with the dynamic spin-spin correlation function
\begin{equation}\begin{aligned}
\label{eq:DynSpinZZCrFt}
S^{zz}(\mathbf{q},\tau) 
&= \frac{1}{N_s}\sum_{\mathbf{ij}} e^{-i\mathbf{q}\cdot(\mathbf{R}_{\mathbf{i}} - \mathbf{R}_{\mathbf{j}})}\times \\
&\hspace{1.0cm} \big(\langle\hat{s}_{\mathbf{i}}^z(\tau)\hat{s}_{\mathbf{j}}^z(0)\rangle-\langle\hat{s}_{\mathbf{i}}^z(\tau)\rangle\langle\hat{s}_{\mathbf{j}}^z(0)\rangle\big),
\end{aligned}\end{equation}
with the spin-$\hat{s}^z$ operator $\hat{s}_{\mathbf{i}}^z=(\hat{n}_{\mathbf{i}\uparrow}-\hat{n}_{\mathbf{i}\downarrow})/2$. We use the static spin susceptibility $S^{zz}(\mathbf{q},i\omega=0)$ to examine the dominant spin fluctuation in the system. 

As demonstrated in Fig.~\ref{fig:U0PairSuscep}, the pairing susceptibilities in singlet and triplet channels are already comparable in noninteracting system. Thus, to investigate the pure correlation effect from the interaction, we also calculate the vertex contributions for both $S(\boldsymbol{\Gamma},i\omega=0)$ and $S^{zz}(\mathbf{q},i\omega=0)$. In DQMC simulations, the vertex contribution for a general correlation function, i.e., $\langle c_1^+c_2^{}c_3^+c_4^{}\rangle$, is computed via subtracting the uncorrelated part~\cite{White1989b,Ying2020}, as $\langle c_1^+c_2^{}c_3^+c_4^{}\rangle - \langle c_1^+c_2^{}\rangle\langle c_3^+c_4^{}\rangle - \langle c_1^+c_4^{}\rangle\langle c_2^{}c_3^+\rangle$ (note that $\langle\cdot\rangle$ denotes the Monte Carlo average).

We note that, in Ref.~\cite{Meng2015}, RPA and cluster DMFT with the Parquet approximation were employed to compute leading eigenvalues of the linearized gap equations for various pairing channels. The most favorable pairing symmetry was then identified as the one associated with the largest leading eigenvalue. However, such calculations involve dynamic two-particle correlation functions, and both methods suffer from uncontrolled approximations, i.e., the mean-field nature of RPA, the locality inherent in the DMFT approximation, and the restricted-level self-consistency used for two-particle correlations in the Parquet approach. {  In comparison, our DQMC calculations of the static pairing susceptibilities and the corresponding vertex contributions across various pairing channels offer clearer and more compelling evidence for a possible spin-triplet instability in the model, via numerically unbiased results in finite-size systems along with an examination of system-size effects.}

% In comparison, our approach employs the determinant quantum Monte Carlo (DQMC) method, which is numerically exact on a finite lattice and does not presuppose any broken symmetries in the simulation itself. We then probe the leading instabilities by evaluating pairing susceptibilities and their vertex contributions across multiple candidate channels on equal footing, albeit limited by the sign problem.

\section{Numerical results}
\label{sec:Results}

In this section, we present our DQMC simulation results and discussions for the model~(\ref{eq:2DHamlt}), including the sign average, pairing susceptibilities in various channels and spin susceptibility with the vertex contributions. Our simulations reach the linear system size $L=16$ and cover interaction strengths $U/t\le 4$, with the temperature as low as $T/t=0.10$. We adopt the discretization step $\Delta\tau t=0.05$, which has been verified to eliminate the Trotter error. Periodic boundary conditions in both directions are applied for all the calculations.

\begin{figure}[t]
\centering
\includegraphics[width=0.88\linewidth]{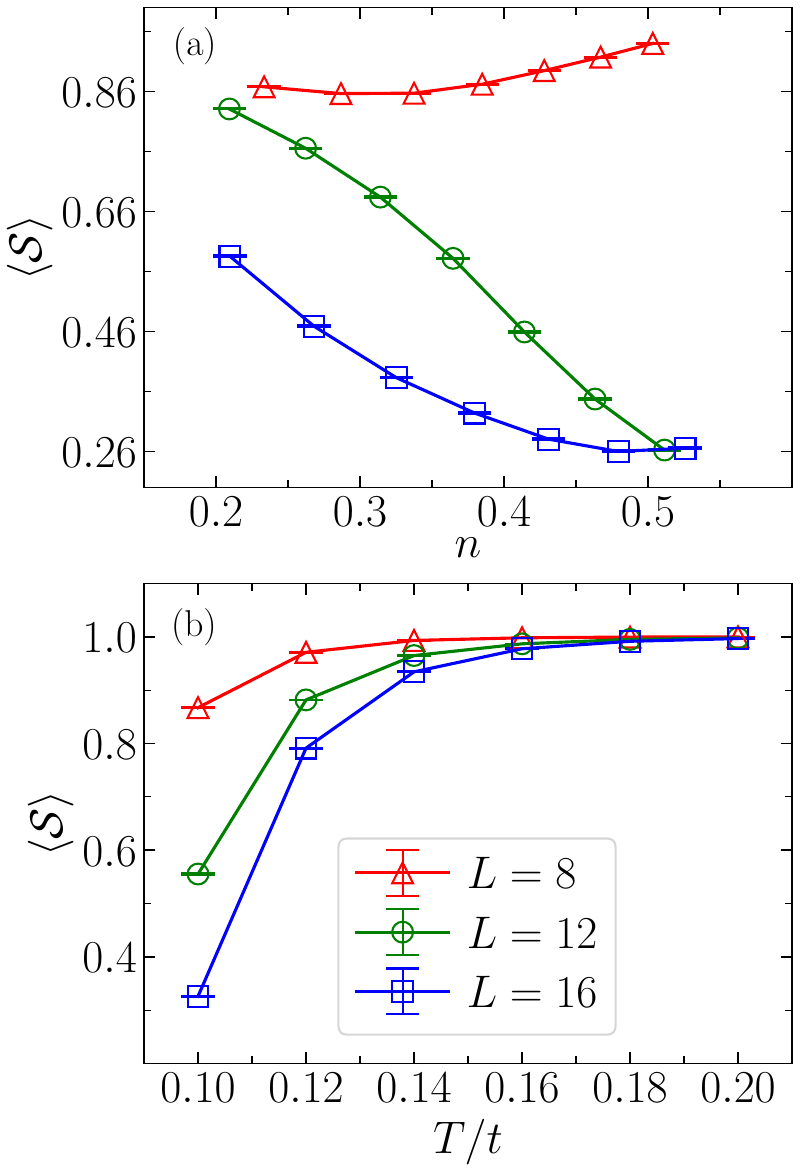}
\caption{Numerical results of the sign average $\langle\mathcal{S}\rangle$ in DQMC simulations for the model~(\ref{eq:2DHamlt}) with $U/t=2$ and $t_2/t=+0.5,t_3/t=-0.1$, from the systems with $L=8,12,16$. Panel (a) plots $\langle\mathcal{S}\rangle$ as a function of electron filling $n$ at $T/t=0.10$, while panel (b) shows $\langle\mathcal{S}\rangle$ versus temperature $T/t$ with fixed $n=n_{\rm v}=0.3757$.}
\label{fig:Sgn}
\end{figure}

\subsection{The sign average}
\label{sec:SignAvg}

The repulsive Hubbard model in Eq.~(\ref{eq:2DHamlt}) suffers from the sign problem in DQMC simulations, due to the presence of $t_2$ and $t_3$ hoppings as well as the hole doping. Employing the reweighting technique discussed in Sec.~\ref{sec:DQMCmethod}, the sign problem typically induces exponentially increasing computational cost versus $\beta$ and $N_s$~\cite{Loh1990,Troyer2005}. 

In Fig.~\ref{fig:Sgn}, we present the numerical results of the sign average $\langle\mathcal{S}\rangle$ as functions of electron filling $n$ and temperature $T/t$, for $U/t=2$ and $t_2/t=+0.5,t_3/t=-0.1$. As shown in Fig.~\ref{fig:Sgn}(a), at $T/t=0.10$ (the lowest temperature accessed in our calculations), $\langle\mathcal{S}\rangle$ shows a pronounced dependence on $n$, and a significant finite-size effect is evident. Specifically, $\langle\mathcal{S}\rangle$ shows no distinct signature associated with the type-II vHs at $n=n_{\rm v}=0.3757$. These observations can be explained by the fact that, in weakly interacting regime, the sign problem is primarily governed by free-fermion spectrum of the model. In the filling range $0.2 \le n \le 0.55$ shown in the plot, the noninteracting system remains in an open-shell structure, i.e., the Fermi level is gapless. In this regime, $\langle\mathcal{S}\rangle$ is influenced by the degeneracy and occupancy of the Fermi energy level in the finite-size system. Furthermore, the free-fermion spectrum depends on the system size, resulting in distinct $n$-dependences for different lattice sizes $L$, which gives rise to the observed finite-size effect. Such a single-particle finite-size effect also exists in the numerical results of pair and spin susceptibilities (see Sec.~\ref{sec:PairSuscep} and Sec.~\ref{sec:SpinSuscep}). By fixing $n=n_{\rm v}$ exactly at the vHs, the results in Fig.~\ref{fig:Sgn}(b) display the characteristic exponential decay of $\langle\mathcal{S}\rangle$ with increasing $\beta$ for all system sizes~\cite{Loh1990,Troyer2005}, with $\langle\mathcal{S}\rangle$$\sim$$1$ for $T/t\ge 0.20$.

As a consequence of the sign problem, our DQMC simulations for the model~(\ref{eq:2DHamlt}) are limited to $L\le 16$ and $T/t\ge 0.10$. In particular, although $\langle\mathcal{S}\rangle$ remains $\sim$$0.30$ at $T/t=0.10$ for the $L=16$ system [see Fig.~\ref{fig:Sgn}(b)], it still leads to substantial statistical fluctuations for the dynamic correlation functions, which typically demands considerably more computational cost to achieve satisfying precision than static observables~\cite{Assaad2008}. Fortunately, the numerical results from such constrained calculations are sufficient to support our findings in this work. {  Furthermore, the system size limitation imposed by the sign problem makes the complete finite-size scaling very challenging, since it requires extrapolating physical observables as a function of $1/L$ to the thermodynamic limit (TDL, $L\to\infty$). Instead, we adopt an alternative finite-size analysis strategy to reach our conclusion for identifying the dominant pairing channel. Specifically, we first quantitatively evaluate the relative strength of all pairing channels using the pairing susceptibility and its vertex contribution, and then examine how these tendencies evolve as the system size increases ($L=8,12,16$) and thereby determining the winning channel(s) in the TDL. This finite-size analysis strategy has been employed in many previous resembling studies~\cite{White1989b,Ying2020,Guo2024} and has been demonstrated to be both efficient and reliable for identifying the dominant pairing channel. }

% As a consequence of the sign problem, our DQMC simulations for the model~(\ref{eq:2DHamlt}) are limited to the system sizes and temperatures shown, up to $L=16$ and down to $T/t=0.10$. While this known limitation precludes a formal finite-size scaling analysis to extrapolate to the thermodynamic limit, the reliability of our conclusions can be assessed differently. At our lowest temperature and largest system size, the average sign $\langle\mathcal{S}\rangle \approx 0.30$ [see Fig.~\ref{fig:Sgn}(b)]. This value, while computationally demanding, is still large enough to obtain statistically robust data for the pairing susceptibilities presented below. Therefore, we base the confidence in our conclusions on the robust and consistent trends observed across all accessible system sizes ($L=8,12,16$). As we will demonstrate in the following sections, the consistent dominance of the $p$-wave pairing channel provides strong evidence that this instability is the primary one in the model and parameter regime studied.

\subsection{Pairing susceptibilities and vertex contributions}
\label{sec:PairSuscep}

In this subsection, we present and discuss the DQMC results of the pairing susceptibilities $S(\boldsymbol{\Gamma},i\omega=0)$ as well as their corresponding vertex contributions in four different pairing channels, including spin-singlet $d_{x^2-y^2}$ and $d_{xy}$ pairings and spin-triplet $p_x$+$ip_y$ and $p_{xy}$ pairings. We focus on the dependences of $S(\boldsymbol{\Gamma},i\omega=0)$ on electron filling $n$, linear system size $L$, temperature $T/t$, and interaction strength $U/t$. For all the presented results, we adopt $t_2/t=-0.5$ and $t_3/t=+0.1$.

\begin{figure}[t]
\centering
\includegraphics[width=0.885\linewidth]{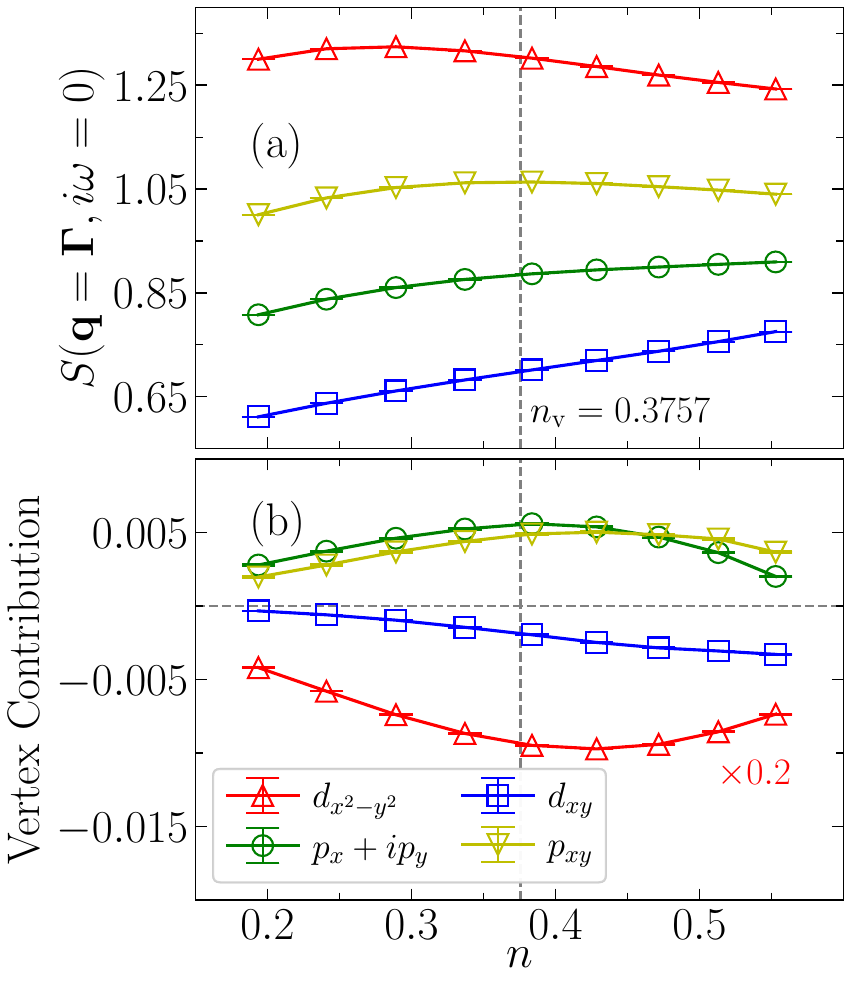}
\caption{DQMC results of (a) pairing susceptibilities $S(\mathbf{q}=\boldsymbol{\Gamma},i\omega=0)$, and (b) their vertex contributions, in $d_{x^2-y^2}$-wave, $d_{xy}$-wave, $p_x$+$ip_y$-wave, and $p_{xy}$-wave pairing channels as a function of electron filling $n$, for the model~(\ref{eq:2DHamlt}) with $U/t=2$ and $t_2/t=+0.5,t_3/t=-0.1$. These results are from calculations for $L=16$ system at $T/t=0.20$. The vertical gray dashed lines in both panels mark the electron filling $n_{\rm v}=0.3757$ at type-II vHs. }
\label{fig:PairSuscep1}
\end{figure}

In Fig.~\ref{fig:PairSuscep1}, we show numerical results versus $n$ around the type-II vHs ($n_{\rm v}=0.3757$), for the model~(\ref{eq:2DHamlt}) with $U/t=2$ at $T/t=0.20$ from $L=16$ system. Based solely on the susceptibility $S(\boldsymbol{\Gamma},i\omega=0)$, the $d_{x^2-y^2}$-wave pairing appears to be the most favorable, as it exhibits the largest value [see Fig.~\ref{fig:PairSuscep1}(a)]. Nevertheless, as shown in Fig.~\ref{fig:PairSuscep1}(b), the vertex contributions for both $d_{x^2-y^2}$-wave and $d_{xy}$-wave pairings are negative, suggesting that the pure interaction effect (also called {\it effective interaction} in Ref.~\cite{White1989b}) actually frustrates these two $d$-wave pairings~\cite{White1989b,Ying2020} in the involved filling region. This also reveals that, for these two pairing channels, the bare results of $S(\boldsymbol{\Gamma},i\omega=0)$ are fully contributed by the uncorrelated part. As a comparison, for $p_x$+$ip_y$-wave and $p_{xy}$-wave pairings, the vertex contributions of $S(\boldsymbol{\Gamma},i\omega=0)$ remain positive. This indicates that the {\it effective interaction} for these two pairing channels is {\it attractive}~\cite{White1989b}, though the uncorrelated part still occupies a large portion in the bare results of $S(\boldsymbol{\Gamma},i\omega=0)$. Moreover, these vertex contributions show broaden peaks around the type-II vHs, which can be attributed to the diverging LDOS at Fermi surface in the noninteracting system. These results provide initial evidence for the possible spin-triplet pairing instability in the lattice model~(\ref{eq:2DHamlt}) near type-II vHs. 

Similar results as those shown in Fig.~\ref{fig:PairSuscep1} are observed at other temperatures in our calculations. The opposite trends observed in the bare susceptibility $S(\boldsymbol{\Gamma},i\omega=0)$ and its vertex contribution for both spin-singlet $d_{x^2-y^2}$-wave and $d_{xy}$-wave pairing channels highlight the critical importance of isolating the pure interaction effects in numerical calculations. In the weak interaction regime, the modest enhancement or suppression of the pairing susceptibility contributed by the {\it effective interaction} can be entirely buried by single-particle contributions originating from the noninteracting part of the model. Such results were also reported in previous studies of pairing instabilities in the doped Hubbard model on honeycomb lattice~\cite{Ying2020,Guo2024}. 

\begin{figure}[t]
\centering
\includegraphics[width=1.00\linewidth]{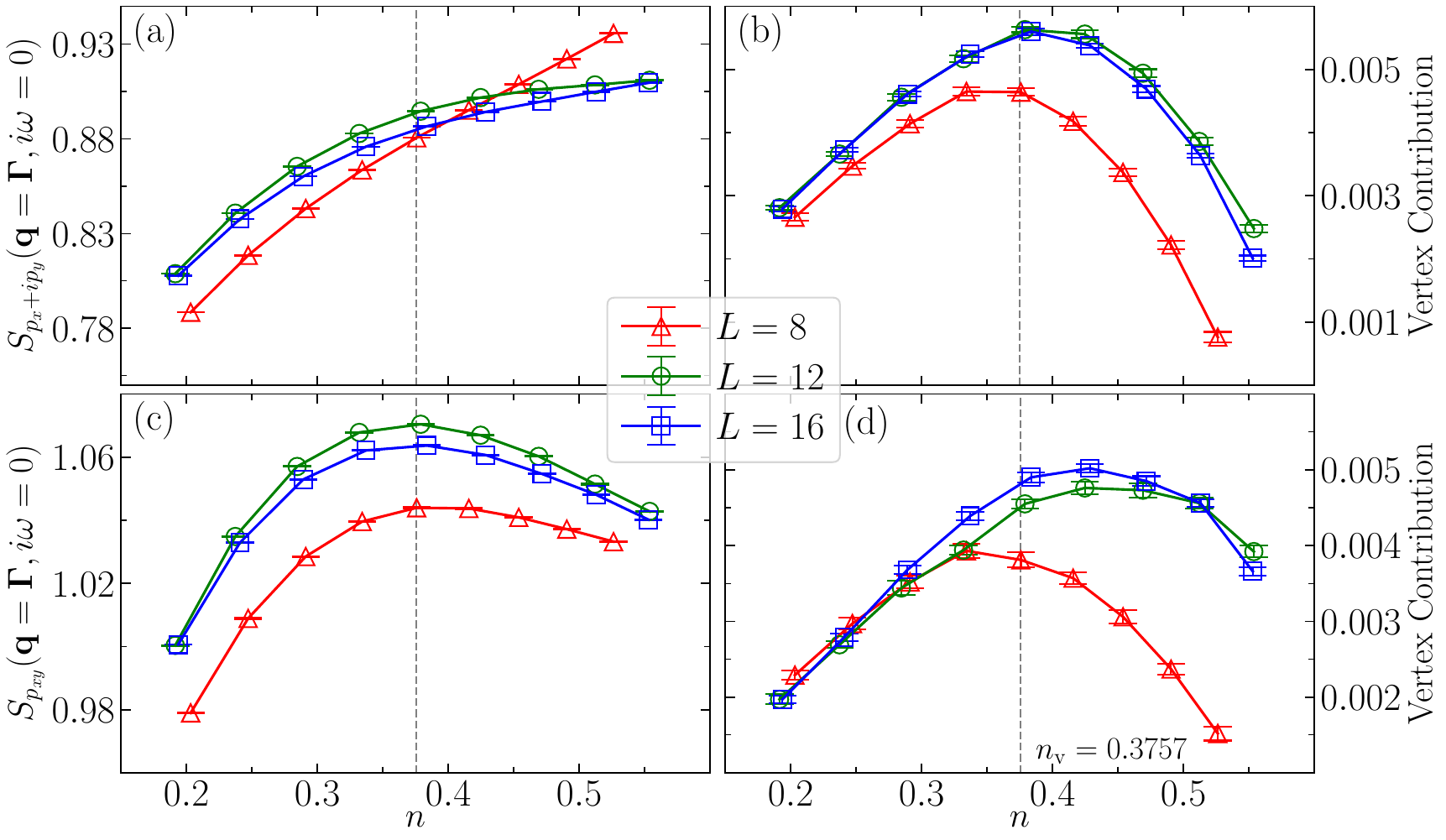}
\caption{DQMC results of the pairing susceptibility and its vertex contribution for (a)(b) the $p_x$+$ip_y$-wave pairing [denoted as $S_{p_x+ip_y}(\boldsymbol{\Gamma},i\omega=0)$], and (c)(d) the $p_{xy}$-wave pairing [denoted as $S_{p_{xy}}(\boldsymbol{\Gamma},i\omega=0)$], versus electron filling $n$, for the model~(\ref{eq:2DHamlt}) with $U/t=2$ and $t_2/t=+0.5,t_3/t=-0.1$. These results are from calculations for the systems with $L=8,12,16$ at $T/t=0.20$. The vertical gray dashed lines mark the electron filling $n_{\rm v}=0.3757$ at type-II vHs. }
\label{fig:PairSuscep2}
\end{figure}

\begin{figure}[t]
\centering
\includegraphics[width=0.935\linewidth]{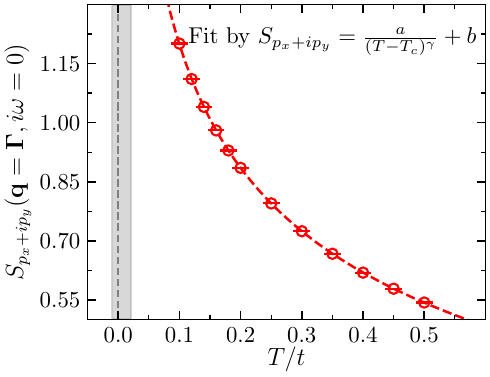}
\caption{Power-law fitting for pairing susceptibility of spin-triplet $p_x$+$ip_y$-wave pairing [denoted as $S_{p_x+ip_y}(\boldsymbol{\Gamma},i\omega=0)$] using the formula $S_{p_x+ip_y}=a(T-T_c)^{-\gamma}+b$, for the model~(\ref{eq:2DHamlt}) with $U/t=2$, $t_2/t=+0.5,t_3/t=-0.1$ and fixed electron filling $n=n_{\rm v}=0.3757$. The red circles are DQMC results from $L=16$ system, and the dashed line represents the fitting curve. The fitting yields $a = 1.4(1)$, $\gamma = -0.21(2)$ and $T_c/t = 0.006(4)$. The gray shaded region marks the estimated $T_c$.}
\label{fig:PairSuscep6}
\end{figure}

Based on the results in Fig.~\ref{fig:PairSuscep1}, we then exam the size dependence of $S(\boldsymbol{\Gamma},i\omega=0)$ and its vertex contribution for the spin-triplet $p_x$+$ip_y$-wave (denoted as $S_{p_x+ip_y}$) and $p_{xy}$-wave (denoted as $S_{p_{xy}}$) pairing channels. The numerical results for $L=8,12$ and $16$ are presented in Fig.~\ref{fig:PairSuscep2}. The bare results of $S_{p_x+ip_y}(\boldsymbol{\Gamma},i\omega=0)$ show enhancements with increasing $n$ in the range $0.20\le n\le 0.55$, whereas those of $S_{p_{xy}}(\boldsymbol{\Gamma},i\omega=0)$ are clearly peaked around $n=n_{\rm v}$. Moreover, both $S_{p_x+ip_y}(\boldsymbol{\Gamma},i\omega=0)$ and $S_{p_{xy}}(\boldsymbol{\Gamma},i\omega=0)$ display oscillating behavior with increasing $L$, which can be attributed to single-particle finite-size effects discussed in Sec.~\ref{sec:SignAvg}. As a comparison, the vertex contributions of both susceptibilities exhibit pronounced peaks near $n=n_{\rm v}$, highlighting the strong attractive {\it effective interaction} for both pairing channels at type-II vHs. The finite-size effect is also rather weak for these vertex contributions, as evidenced by the convergence for the $p_x$+$ip_y$ channel at $L=16$, and the slight increase for the $p_{xy}$ channel from $L=12$ to $L=16$. These weak size dependences for both $S_{p_x+ip_y}(\boldsymbol{\Gamma},i\omega=0)$ and $S_{p_{xy}}(\boldsymbol{\Gamma},i\omega=0)$ as well as their vertex contributions might indicate that, the temperature $T/t=0.20$ for the results in Fig.~\ref{fig:PairSuscep1}, is relatively high comparing to the possible superconducting transition of spin-triplet pairing in the system. This point is confirmed by our results of estimating the transition temperature (see Fig.~\ref{fig:PairSuscep6} and the corresponding discussion).

The results shown in Figs.~\ref{fig:PairSuscep1} and \ref{fig:PairSuscep2} provide preliminary evidence that, the system favors spin-triplet $p_x$+$ip_y$-wave and $p_{xy}$-wave pairings, with the strongest tendency appearing near the electron filling $n_{\rm v}$ corresponding to the type-II vHs. Moreover, the vertex contributions explicitly illustrate that these two spin-triplet pairing channels are almost equally favorable. This behavior is reasonable from the perspective of $p$-wave symmetry, which allows the coexistence of these two pairing orders. A similar phenomenon was also observed in the mixed-parity pairing in the 2D attractive Hubbard model with spin-orbit coupling~\cite{Song2024}, where the spin-triplet pairing component primarily consists of NN and NNN pairings in real space. Their relative strengths depend on both the interaction strength and the electron filling, and can be tuned to be comparable~\cite{Song2024}. {  Importantly, as discussed in Sec.~\ref{sec:SignAvg}, although a formal finite-size scaling extrapolation is not feasible due to the system size limitation imposed by the sign problem, the dominance of the $p$-wave channels persists consistently across all accessible system sizes ($L=8,12,16$), thereby providing strong support for the drawn conclusions.} Based on these insights, in the following, we fix the electron filling to $n=n_{\rm v}$, and investigate the dependences of pairing susceptibilities on temperature and interaction strength. 

\begin{figure}[t]
\centering
\includegraphics[width=0.84\linewidth]{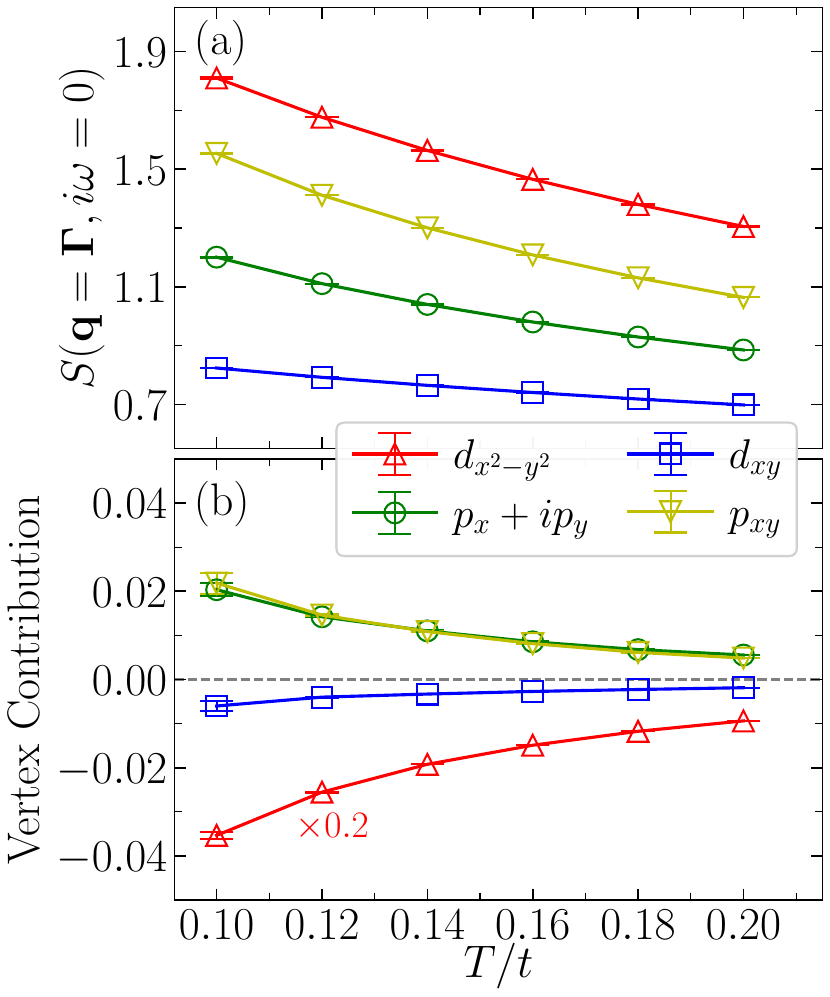}
\caption{DQMC results of (a) pairing susceptibilities $S(\mathbf{q}=\boldsymbol{\Gamma},i\omega=0)$, and (b) their vertex contributions, in $d_{x^2-y^2}$-wave, $d_{xy}$-wave, $p_x$+$ip_y$-wave, and $p_{xy}$-wave pairing channels as a function of temperature $T/t$ at fixed $n=n_{\rm v}=0.3757$, for the model~(\ref{eq:2DHamlt}) with $U/t=2$ and $t_2/t=+0.5,t_3/t=-0.1$. These results are from calculations for $L=16$ system.}
\label{fig:PairSuscep3}
\end{figure}

In Fig.~\ref{fig:PairSuscep3}, we present the temperature dependence of $S(\boldsymbol{\Gamma},i\omega=0)$ and the vertex contribution for all four pairing channels, in the range $0.10\le T/t\le 0.20$, for $U/t=2$ from $L=16$ system at $n=n_{\rm v}$. As shown in Fig.~\ref{fig:PairSuscep3}(a), all $S(\boldsymbol{\Gamma},i\omega=0)$ values increase as temperature decreases, with the spin-singlet $d_{x^2-y^2}$-wave pairing exhibiting the largest magnitude across the entire temperature range. This trend is the same as that observed in the noninteracting system as shown in Fig.~\ref{fig:U0PairSuscep}. However, as illustrated in Fig.~\ref{fig:PairSuscep3}(b), the vertex contributions for both $d_{x^2-y^2}$-wave and $d_{xy}$-wave pairings remain consistently negative, with their absolute values increasing monotonically as the temperature decreases. This behavior indicates that the repulsive {\it effective interaction} for these two spin-singlet pairing channels becomes even stronger upon cooling. These results basically exclude the spin-singlet $d_{x^2-y^2}$-wave and $d_{xy}$-wave pairing instabilities in the system at type-II vHs. Conversely, the vertex contributions for $p_x$+$ip_y$-wave and $p_{xy}$-wave pairings are always positive and exhibit steady enhancements as the temperature decreases [see Fig.~\ref{fig:PairSuscep3}(b)]. This clearly demonstrates that the spin-triplet pairing instability at type-II vHs becomes increasingly pronounced towards low temperature region. Moreover, vertex contributions for the two triplet channels are nearly degenerate in the plotted temperature range, suggesting the coexistence feature of these two pairing orders. 

\begin{figure}[t]
\centering
\includegraphics[width=0.99\linewidth]{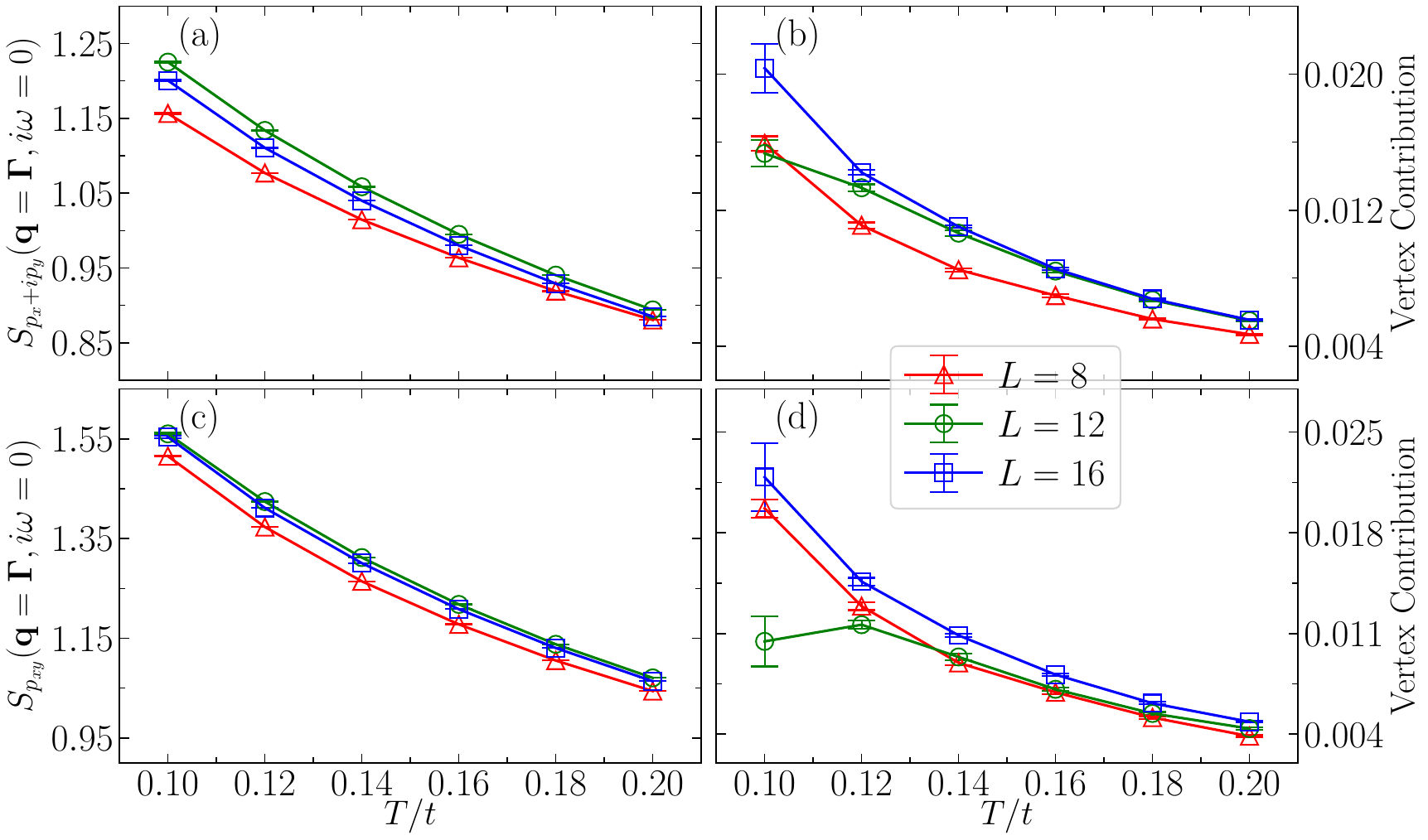}
\caption{DQMC results of the pairing susceptibility and its vertex contribution for (a)(b) the $p_x$+$ip_y$-wave pairing [denoted as $S_{p_x+ip_y}(\boldsymbol{\Gamma},i\omega=0)$], and (c)(d) the $p_{xy}$-wave pairing [denoted as $S_{p_{xy}}(\boldsymbol{\Gamma},i\omega=0)$], as a function of temperature $T/t$, for the model~(\ref{eq:2DHamlt}) with $U/t=2$, $t_2/t=+0.5,t_3/t=-0.1$ and fixed electron filling $n=n_{\rm v}=0.3757$. These results are from calculations for the systems with $L=8,12,16$. }
\label{fig:PairSuscep4}
\end{figure}

In line with Fig.~\ref{fig:PairSuscep3}, we accordingly show the size dependence of $S_{p_x+ip_y}(\boldsymbol{\Gamma},i\omega=0)$ and $S_{p_{xy}}(\boldsymbol{\Gamma},i\omega=0)$ as well as their vertex contributions versus temperature in Fig.~\ref{fig:PairSuscep4}. Both the bare susceptibilities and vertex contributions increase with decreasing temperature for all system sizes. Once again, the single-particle finite-size effect induces the oscillatory behaviors of size dependence, with the $L=16$ results lying between those of $L=8$ and $L=12$. For $L=12$ at $T/t\le 0.14$, the abnormal results in vertex contributions of both $S_{p_x+ip_y}(\boldsymbol{\Gamma},i\omega=0)$ and $S_{p_{xy}}(\boldsymbol{\Gamma},i\omega=0)$ might share the same origin. Another notable feature is that the enhancement of the vertex contributions becomes more rapid at lower temperatures (except for $L=12$). Combining the results from Fig.~\ref{fig:PairSuscep3} and Fig.~\ref{fig:PairSuscep4}, it is highly likely that spin-triplet pairing instability dominates the low-temperature property of the system, potentially leading to a stable ground state characterized by long-range spin-triplet pairing orders.

\begin{figure}[t]
\centering
\includegraphics[width=0.880\linewidth]{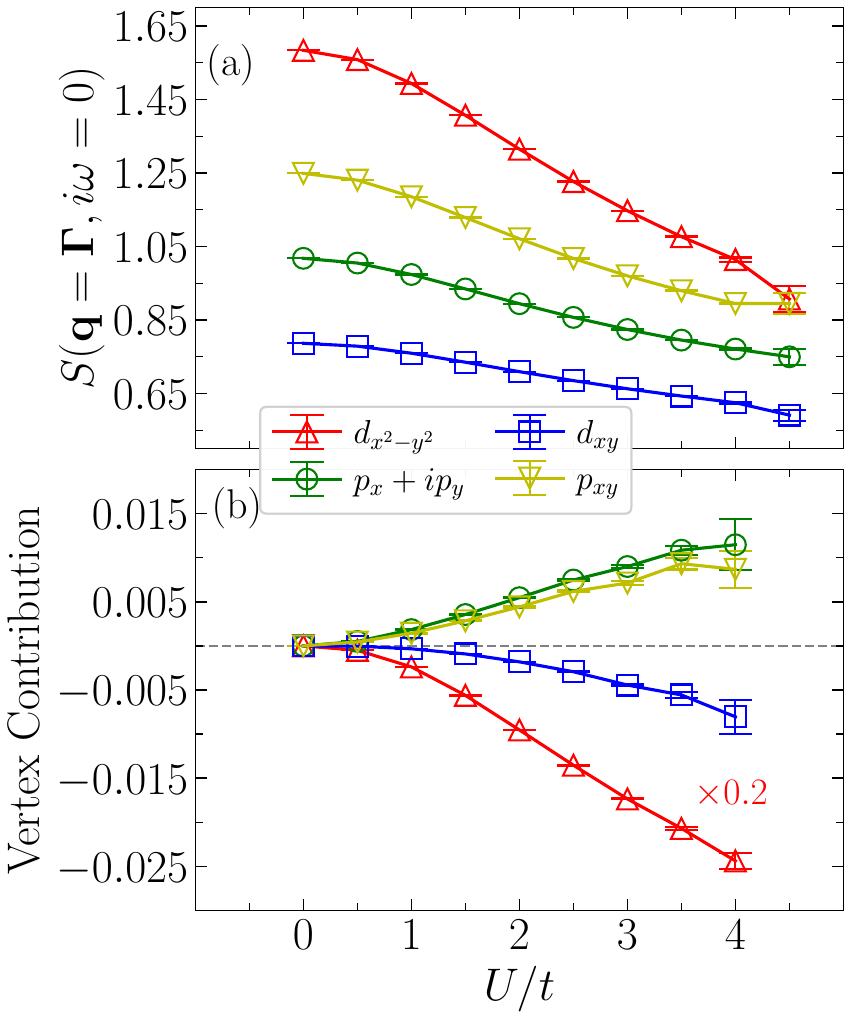}
\caption{DQMC results of (a) pairing susceptibilities $S(\mathbf{q}=\boldsymbol{\Gamma},i\omega=0)$, and (b) their vertex contributions, in $d_{x^2-y^2}$-wave, $d_{xy}$-wave, $p_x$+$ip_y$-wave, and $p_{xy}$-wave pairing channels as a function of interaction strength $U/t$, for the model~(\ref{eq:2DHamlt}) with $t_2/t=+0.5,t_3/t=-0.1$  and fixed electron filling $n=n_{\rm v}=0.3757$. These results are from calculations for $L=12$ system at $T/t=0.20$. }
\label{fig:PairSuscep5}
\end{figure}

In the preceding numerical results, we concentrate on the interaction strength $U/t=2$. Now we proceed to study the evolution of pairing susceptibilities and vertex contributions with increasing $U/t$. In Fig.~\ref{fig:PairSuscep5}, we present the results in the range $0\le U/t\le 4$ for $L=12$ at $T/t=0.20$ for the model at fixed filling $n=n_{\rm v}$. We find that the bare susceptibilities for all four pairing channels are suppressed as $U/t$ increases [see Fig.~\ref{fig:PairSuscep5}(a)]. For the spin-singlet $d_{x^2-y^2}$-wave and $d_{xy}$-wave pairings, the vertex contributions remain negative and grow in magnitude with increasing $U/t$ [see Fig.~\ref{fig:PairSuscep5}(b)], indicating that these spin-singlet channels become even less favorable at stronger interactions. In contrast, the vertex contributions for spin-triplet $p_x$+$ip_y$-wave and $p_{xy}$-wave pairings consistently increase with $U/t$, revealing that the {\rm effective interaction} becomes stronger in these channels. However, regarding the opposite trends observed between $S(\mathbf{q}=\boldsymbol{\Gamma},i\omega=0)$ and the vertex contribution for these two triplet channels, it remains unclear whether the spin-triplet instability persists at larger $U/t$. This uncertainty is further compounded by the limitations of our DQMC simulations, which are restricted to $U/t\le 4$ due to the sign problem. These results confirm that the spin-triplet instability in the model at type-II vHs can sustain in the weak to intermediate interaction regime. 

Based on the above results, we conclude that spin-triplet pairing constitutes the leading instability of the model~(\ref{eq:2DHamlt}) near the type-II vHs at low temperatures. This implies a superconducting ground state characterized by long-range spin-triplet pairing orders ($p_x$+$ip_y$-wave and $p_{xy}$-wave) for the model. Consequently, a superfluid transition of Berezinskii-Kosterlitz-Thouless type is expected to occur at a finite critical temperature, where the corresponding pairing susceptibility diverges. Using this diverging feature, we apply a fitting for $S_{p_x+ip_y}(\boldsymbol{\Gamma},i\omega=0)$ versus temperature with the formula $S_{p_x+ip_y}=a(T-T_c)^{-\gamma}+b$, where $T_c$ is the superfluid transition temperature. The fitting formula can be taken as a modified Curie-Weiss law for the susceptibility in normal phase. A similar fitting procedure was applied in the experimental observation of the antiferromagnetic phase transition in three-dimensional Hubbard model in optical lattice~\cite{Shao2024}. Moreover, as shown in Fig.~\ref{fig:PairSuscep4}, the finite-size effect is rather weak in $S_{p_x+ip_y}(\boldsymbol{\Gamma},i\omega=0)$, and thus we perform the fitting for $L=16$ data. We present the fitting result in Fig.~\ref{fig:PairSuscep6} for $U/t=2$ at the electron filling $n=n_{\rm v}$ (exactly at type-II vHs). The fitting procedure yields a very low transition temperature of $T_c/t=0.006(4)$ can be reached from the fitting procedure, which is well below the accessible range of our current DQMC simulations. A more reliable investigation of the potential spin-triplet superfluid transition in the model~(\ref{eq:2DHamlt}) near type-II vHs would require an effective strategy to overcome the sign problem, such as employing the constrained-path quantum Monte Carlo method~\cite{Yuanyao2019,Zhang1999,Xiao2023}. We leave such an investigation for future work.

\subsection{Spin susceptibility and vertex contribution}
\label{sec:SpinSuscep}

In previous studies, it was revealed that the spin-triplet pairing is typically related to the ferromagnetic spin fluctuations~\cite{Anderson1973,Maeno2024,Sheng2019,Avers2020,Meng2015}. To verify this relation, we compute the static spin susceptibility $S^{zz}(\mathbf{q},i\omega=0)$ with Eq.~(\ref{eq:DynSpinZZCrFt}) in our DQMC simulations for the model~(\ref{eq:2DHamlt}).

\begin{figure}[t]
\centering
\includegraphics[width=0.945\linewidth]{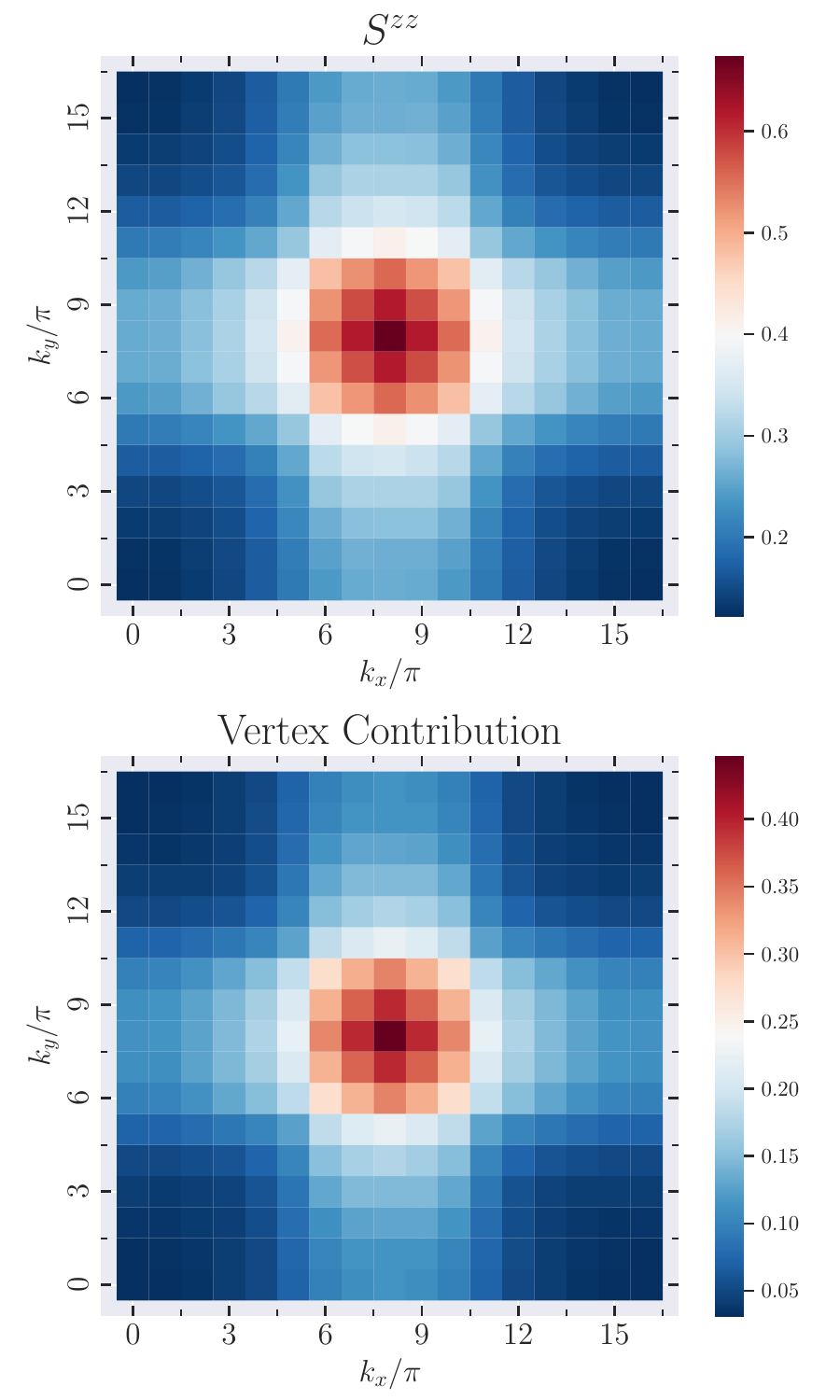}
\caption{Contour plots of (a) spin susceptibility $S^{zz}(\mathbf{q},i\omega=0)$, and (b) its vertex contribution, for the model~(\ref{eq:2DHamlt}) with $U/t=2$, $t_2/t=+0.5,t_3/t=-0.1$ and fixed electron filling $n=n_{\rm v}=0.3757$. These results are from calculations for the system with $L=16$ at $T/t=0.10$.}
\label{fig:L16_Szz}
\end{figure}

\begin{figure}[t]
\centering
\includegraphics[width=0.932\linewidth]{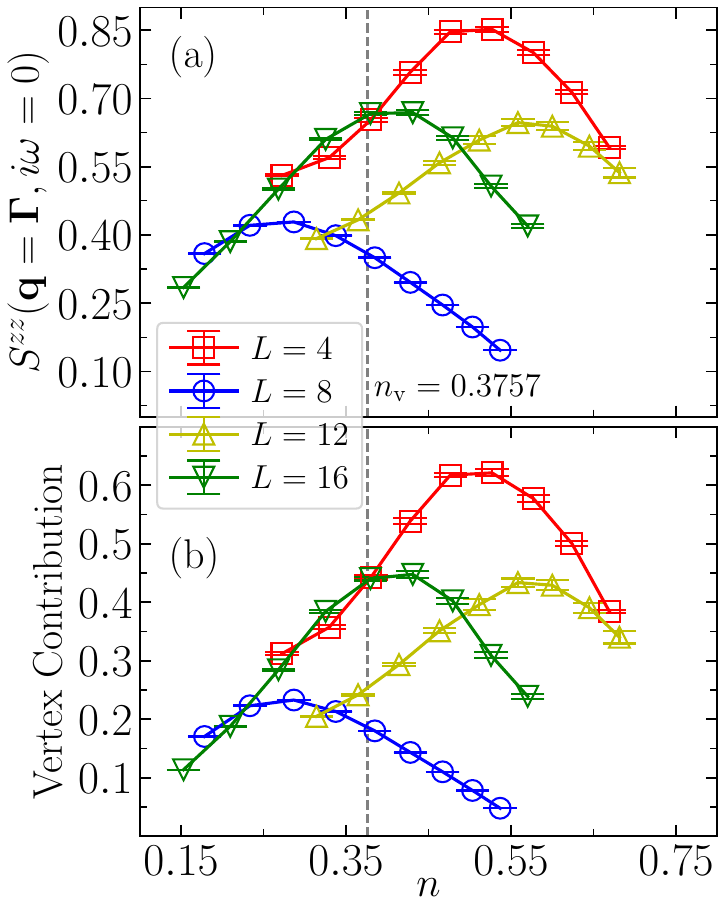}
\caption{DQMC results of (a) spin susceptibility $S^{zz}(\mathbf{q},i\omega=0)$, and (b) its vertex contribution, as a function of electron filling $n$, for the model~(\ref{eq:2DHamlt}) with $U/t=2$ and $t_2/t=+0.5,t_3/t=-0.1$. These results are from calculations for the systems with $L=4,8,12,16$ at $T/t=0.10$. The vertical gray dashed lines in (a) and (b) mark the electron filling $n_{\rm v}=0.3757$ at type-II vHs. }
\label{fig:Szz_rho}
\end{figure}

In Fig.~\ref{fig:L16_Szz}, we present the contour plots of DQMC results for $S^{zz}(\mathbf{q},i\omega=0)$ and its vertex contribution versus $k_x$ and $k_y$ on a $16 \times 16$ lattice at $T/t=0.10$. The data are obtained for the model~(\ref{eq:2DHamlt}) with $U/t = 2$ and $t_2/t=+0.5,t_3/t=-0.1$ at electron filling $n = n_{\rm v} = 0.3757$. It is evident that both quantities show prominent peaks at $\boldsymbol{\Gamma}=(0,0)$ point, suggesting the dominant ferromagnetic spin correlations in the system at type-II vHs. Moreover, the values of vertex contribution around $\boldsymbol{\Gamma}$ point occupy more than $60\%$ of the bare $S^{zz}(\mathbf{q},i\omega=0)$ results, which illustrates the significant contribution from the pure interaction effect to the ferromagnetic spin fluctuation. This is in accordance with the {\rm effective interaction} of attractive type for spin-triplet $p_x$+$ip_y$-wave and $p_{xy}$-wave pairings in the model. 

To fully characterize the relationship between spin and pairing instabilities, we vary the electron filling around type-II vHs, and compute both the ferromagnetic spin susceptibility $S^{zz}(\mathbf{q}=\boldsymbol{\Gamma},i\omega=0)$ and its vertex contribution. The corresponding numerical results are presented in Fig.~\ref{fig:Szz_rho}. Pronounced finite-size effects are evident in both quantities, as the results from different system sizes exhibit peaks at varying values of $n$. This behavior can be largely attributed to single-particle finite-size effects as discussed in Sec.~\ref{sec:SignAvg} and Sec.~\ref{sec:SpinSuscep}. Nevertheless, as the linear system size $L$ increases, the peak positions of both $S^{zz}(\boldsymbol{\Gamma},i\omega=0)$ and its vertex part show an oscillating convergence toward $n = n_{\rm v}$. Notably, the vertex contributions from the $L=16$ system display peak structures that closely resemble those of the spin-triplet $p_x+ip_y$-wave and $p_{xy}$-wave pairing susceptibilities shown in Fig.\ref{fig:PairSuscep2}. This again strongly suggests the intrinsic connection between the ferromagnetic spin fluctuation and the spin-triplet pairing instability in the model. 

The numerical results in Figs.~\ref{fig:L16_Szz} and~\ref{fig:Szz_rho} explicitly show that the leading spin fluctuations in the model~(\ref{eq:2DHamlt}) around type-II vHs is of ferromagnetic type. These results also offer reasonable explanations for the results of various pairing susceptibilities presented and discussed in Sec.~\ref{sec:SpinSuscep}. Since the spin-singlet pairing is typically accompanied by antiferromagnetic spin fluctuation~\cite{Dagotto1994,Tsuei2000}, the $d_{x^2-y^2}$-wave and $d_{xy}$-wave pairings are naturally frustrated by the ferromagnetic correlations. On the contrary, the ferromagnetic spin fluctuations in the model~(\ref{eq:2DHamlt}) around type-II vHs can mediate and stabilize the spin-triplet pairings in this repulsive system.

\section{Summary and discussion}
\label{sec:Summary}

To summarize, we investigate the pairing instabilities in various symmetry channels in a square-lattice Hubbard model near type-II vHs, using numerically exact DQMC method. Based on the pairing susceptibility and its vertex contribution, our numerical results clearly reveal the spin-triplet pairing instabilities in $p_x$+$ip_y$-wave and $p_{xy}$-wave pairing channels towards the low-temperature region. We also illustrate that the ferromagnetic spin fluctuations in the model is very likely to serve as the mechanism for the spin-triplet superconductivity. {  Our results are qualitatively consistent with previous studies based on the RG analysis~\cite{Yao2015} and DMFT simulation~\cite{Meng2015}, while offering an unbiased numerical perspective that is complementary to these approximate methods. }

%Our approach employs DQMC, which is numerically exact on a finite lattice and does not presuppose any broken symmetries in the simulation; leading tendencies are identified by post-processing correlation functions and susceptibilities across candidate channels.

Our work has implications with the realizations of spin-triplet superconductivity and superfluidity. In previous studies~\cite{Chenyue2022,Han2023}, it was proposed that, the spin-triplet $p$-wave pairing can emerge in the repulsive Hubbard model with opposite hopping signs in spin-up and -down sectors. This is reasonable since the spin imbalance naturally induces ferromagnetic spin fluctuations that can stabilize the triplet pairing. However, our study focus on the spin-balanced Hubbard model, and the ferromagnetic spin fluctuations appears due to the type-II vHs. Therefore, our work should be taken as a more meaningful approach to realize the spin-triplet superconductivity in the repulsive Hubbard model. On the other hand, the most recent optical lattice experiment using ultracold atoms~\cite{Xu2025} has achieved the low temperature $\sim$$0.05t$ (with $t$ as the nearest-neighbor hopping strength) for 2D Hubbard model. This opens up the possibility of verifying our numerical results and observing both ferromagnetic spin correlations and spin-triplet pairing instability in the model~(\ref{eq:2DHamlt}) near the type-II vHs.

\begin{acknowledgments}
    This work was supported by the Natural Science Foundation of Heilongjiang Province (under Grant No.YQ2023A004), the National Natural Science Foundation of China (under Grants No.12247103 and No.12204377), the Quantum Science and Technology-National Science and Technology Major Project (Grant No.2021ZD0301900), and the Youth Innovation Team of Shaanxi Universities.
\end{acknowledgments}

\appendix
\section{Pairing Susceptibility in a Non-Interacting System}
\label{sec:AppendixA}

The dynamical pairing correlation function is defined as:
\begin{equation}
\begin{split}
    S(\tau,{\mathbf{q}}) &= \frac{1}{L^2}\sum_{{\mathbf{i}},{\mathbf{j}}} e^{-i{\mathbf{q}}\cdot({\mathbf{R}}_i - {\mathbf{R}}_j)} 
    \langle \hat{\Delta}_{\mathbf{i}}(\tau) \hat{\Delta}_{\mathbf{j}}(0) \rangle \\
    &= \langle e^{\tau \hat{H}} \hat{\Delta}_{\mathbf{q}} e^{-\tau \hat{H}} \hat{\Delta}_{-{\mathbf{q}}} \rangle
\end{split}
\label{eq:A1}
\end{equation}
Here, $\tau$ denotes the imaginary time, ${\mathbf{q}}$ is the pairing momentum, and $\hat{\Delta}_{\mathbf{q}} = L^{-1}\sum_{\mathbf{i}} \hat{\Delta}_{\mathbf{i}} e^{-i{\mathbf{q}}\cdot{\mathbf{R}}_i}$ is the Fourier-transformed pair creation operator. The explicit form of this operator in momentum space is:
\begin{equation}
    {{\hat \Delta }_{\bf{q}}} = \frac{1}{L}\sum\limits_{\bf{k}} {\left( {f({\bf{k}}){{\hat P}_{{\bf{k}},{\bf{q}}}} + f{{({\bf{k}})}^*}\hat P_{{\bf{k}}, - {\bf{q}}}^\dag } \right)}
\label{eq:A2}
\end{equation}
where ${{\hat P}_{{\bf{k}},{\bf{q}}}} \equiv {{\hat c}_{ - {\bf{k}} + {\bf{q}} \uparrow }}{{\hat c}_{{\bf{k}} \downarrow }} \pm {{\hat c}_{ - {\bf{k}} + {\bf{q}} \downarrow }}{{\hat c}_{{\bf{k}} \uparrow }}$ is the pair annihilation operator. The upper ($+$) and lower ($-$) signs correspond to spin-triplet and spin-singlet pairing, respectively.

For the non-interacting Hamiltonian $\hat{H}_0 = \sum\nolimits_{{\bf{ij}},\sigma } {{t_{{\bf{ij}},\sigma }}\hat{c}_{{\bf{i}}\sigma }^ \dagger {\hat{c}_{{\bf{j}}\sigma }}}  = \sum\nolimits_{{\bf{k}}\sigma } {{\varepsilon _{{\bf{k}}\sigma }}{\hat{c}_{{\bf{k}}\sigma }}{\hat{c}_{{\bf{k}}\sigma }}}$, the dynamical Green's functions are:
\begin{equation}
    \begin{matrix}
        \left\langle \hat{c}_{{\mathbf{k}}\sigma}(\tau) \hat{c}_{{\mathbf{k}}\sigma}^{\dagger}(0) \right\rangle = e^{-\tau \varepsilon_{{\mathbf{k}}\sigma}} \bar{f}_{{\mathbf{k}}\sigma} \\[3pt]
        \left\langle \hat{c}_{{\mathbf{k}}\sigma}(0) \hat{c}_{{\mathbf{k}}\sigma}^{\dagger}(\tau) \right\rangle = e^{+\tau \varepsilon_{{\mathbf{k}}\sigma}} \bar{f}_{{\mathbf{k}}\sigma} \\[3pt]
        \left\langle \hat{c}_{{\mathbf{k}}\sigma}^{\dagger}(0) \hat{c}_{{\mathbf{k}}\sigma}(\tau) \right\rangle = e^{-\tau \varepsilon_{{\mathbf{k}}\sigma}} f_{{\mathbf{k}}\sigma} \\[3pt]
        \left\langle \hat{c}_{{\mathbf{k}}\sigma}^{\dagger}(\tau) \hat{c}_{{\mathbf{k}}\sigma}(0) \right\rangle = e^{+\tau \varepsilon_{{\mathbf{k}}\sigma}} f_{{\mathbf{k}}\sigma}
    \end{matrix}
\label{eq:A3}
\end{equation}
with $f_{{\mathbf{k}}\sigma} = (e^{\beta \varepsilon_{{\mathbf{k}}\sigma}} + 1)^{-1}$ being the Fermi-Dirac distribution, $\bar{f}_{{\mathbf{k}}\sigma} = 1 - f_{{\mathbf{k}}\sigma}$, and $\beta$ the inverse temperature.
Using Wick's theorem for the non-interacting system and the Green's functions in Eq.~(\ref{eq:A3}), the pairing correlation function evaluates to:
\begin{equation}
\begin{aligned}
S(\tau, {\mathbf{q}}) = \frac{1}{L^2} \sum_{{\mathbf{k}}} \Bigl[ &
A^{+}({\mathbf{k}}, {\mathbf{q}}) T_{-}({\mathbf{k}}, {\mathbf{q}}) \\
+ &A^{-}({\mathbf{k}}, {\mathbf{q}}) T_{+}({\mathbf{k}}, {\mathbf{q}}) \Bigr],
\end{aligned}
\label{eq:A4}
\end{equation}
where $A^{\pm}({\mathbf{k}}, {\mathbf{q}})$ and $T_{\pm}({\mathbf{k}}, {\mathbf{q}})$ are defined as follows.
The pairing symmetry factors are:
\begin{equation}
\begin{aligned}
{A^ + }({\bf{k}},{\bf{q}}) \equiv f({\bf{k}})\left[ {f{{({\bf{k}})}^*} \mp f{{( - {\bf{k}} + {\bf{q}})}^*}} \right] \\
{A^ - }({\bf{k}},{\bf{q}}) \equiv f{({\bf{k}})^*}\left[ {f({\bf{k}}) \mp f( - {\bf{k}} - {\bf{q}})} \right]
\end{aligned}
\label{eq:A5}
\end{equation}
preserving the original definition of $f({\mathbf{k}})$ from Eq.~(\ref{eq:A2}).
The weight factors are:
\begin{equation}
\begin{aligned}
T_-(\mathbf{k}, \mathbf{q}) &\equiv 
e^{-\tau (\varepsilon_{-\mathbf{k}+\mathbf{q}\uparrow} + \varepsilon_{\mathbf{k}\downarrow})}
\bar{f}_{\mathbf{k}\downarrow}
\bar{f}_{-\mathbf{k}+\mathbf{q}\uparrow} \\
&\quad + 
e^{-\tau (\varepsilon_{-\mathbf{k}+\mathbf{q}\downarrow} + \varepsilon_{\mathbf{k}\uparrow})}
\bar{f}_{\mathbf{k}\uparrow}
\bar{f}_{-\mathbf{k}+\mathbf{q}\downarrow} \\
T_+(\mathbf{k}, \mathbf{q}) &\equiv 
e^{+\tau (\varepsilon_{\mathbf{k}\downarrow} + \varepsilon_{-\mathbf{k}-\mathbf{q}\uparrow})}
f_{\mathbf{k}\downarrow}
f_{-\mathbf{k}-\mathbf{q}\uparrow} \\
&\quad + 
e^{+\tau (\varepsilon_{\mathbf{k}\uparrow} + \varepsilon_{-\mathbf{k}-\mathbf{q}\downarrow})}
f_{\mathbf{k}\uparrow}
f_{-\mathbf{k}-\mathbf{q}\downarrow}
\end{aligned}
\label{eq:A6}
\end{equation}
The corresponding Matsubara frequency susceptibility $S(\mathbf{q},i\omega_n)$ is obtained via $\int_0^\beta d\tau e^{i\omega_n \tau} S(\tau,\mathbf{q})$, with $\omega_n = 2n\pi/\beta$ being bosonic Matsubara frequencies.

\bibliography{vhsmain}
\end{document}